\documentclass[twocolumn,epsfig,superscriptaddress,prl]{revtex4-2}

\usepackage{graphicx}
\usepackage{amssymb}
\usepackage{amsmath}
\usepackage{hyperref}
\usepackage{changes}
\usepackage{comment}
\usepackage{mathtools}
\usepackage[caption=false]{subfig} 
\usepackage{ragged2e} 
\DeclareCaptionJustification{justified}{\justifying}
\usepackage{cancel}
\usepackage{stackengine}
\usepackage{placeins}
\usepackage[percent]{overpic}
\usepackage{tikz}
\usetikzlibrary{arrows.meta}

\DeclareMathAlphabet{\mathpzc}{OT1}{pzc}{m}{it}

\begin{document}
\title{Nonequilibrium Fractional Josephson Effect}	

\author{Aritra Lahiri}
\email[]{aritra.lahiri@physik.uni-wuerzburg.de}
\affiliation{Institute for Theoretical Physics and Astrophysics,
University of W\"urzburg, D-97074 W\"urzburg, Germany}
\author{Sang-Jun Choi}
\email[]{sang-jun.choi@physik.uni-wuerzburg.de}
\affiliation{Institute for Theoretical Physics and Astrophysics, University of W\"urzburg, D-97074 W\"urzburg, Germany}
\author{Björn Trauzettel}
\affiliation{Institute for Theoretical Physics and Astrophysics, University of W\"urzburg, D-97074 W\"urzburg, Germany}
\affiliation{W\"urzburg-Dresden Cluster of Excellence ct.qmat, Germany}
\date{\today}

\begin{abstract}
{Josephson tunnel junctions exhibit a supercurrent typically proportional to the sine of the superconducting phase difference $\phi$. In general, a term proportional to $\cos(\phi)$ is also present, alongside microscopic electronic retardation effects. We show that voltage pulses sharply varying in time prompt a significant impact of the $\cos(\phi)$ term. Its interplay with the $\sin(\phi)$ term results in a nonequilibrium fractional Josephson effect (NFJE) $\sim\sin(\phi/2)$ in the presence of bound states close to zero frequency. Our microscopic analysis reveals that the interference of nonequilibrium virtual quasiparticle excitations is responsible for this phenomenon. We also analyse this phenomenon for topological Josephson junctions with Majorana bound states. Remarkably, the NFJE is independent of the ground state fermion parity unlike its equilibrium counterpart.}
\end{abstract}
\maketitle

The Josephson effect~\cite{Josephson1962,Josephson1964,Josephson1965,Anderson1963,Rowell1963,Yanson1965}, characterised by coherent tunneling of Cooper pairs across a superconducting junction, is the quintessential manifestation of superconducting phase coherence. In its simplest version, it produces an equilibrium supercurrent with the current-phase relation (C$\phi$R) $I_S(\phi)\sim\sin(\phi)$, where $\phi(t)$ is the superconducting phase difference satisfying $d\phi(t)/dt=2eV(t)/\hbar$, equaling the Josephson frequency $\omega_J=2eV/\hbar$ for constant bias. Typically, time-reversal symmetry implies $I_S(\phi)=-I_S(-\phi)$ which, combined with the periodicity $I_S(\phi+2\pi)=I_S(\phi)$, prompts a sinusoid as the basic model for Josephson \emph{tunnel} junctions~\cite{Golubov2004}. However, microscopic descriptions reveal that the C$\phi$R is altered by the retardation imparted by intermediate electronic excitations~\cite{Josephson1962,Ambegaokar1963,Werthamer1966,Larkin1967,Choi2022,Barone1982}. Specifically, the current equals $I[\phi(t)]=I_N[\phi(t)]+I_S[\phi(t)]$, with $I_N[\phi(t)]=\Re J_{n}[\phi(t)]$ and $I_S[\phi(t)]=\Re J_p[\phi(t)]\sin(\phi(t))-\Im J_p[\phi(t)]\cos(\phi(t))$, where $J_{n,p}[\phi(t)]$ characterise the electronic retardation~\cite{Josephson1962,Ambegaokar1963,Werthamer1966,Larkin1967,Choi2022,Harris1974}. $I_N$ describes the resistive quasi-particle current~\cite{Giaever1960a,Giaever1960b,Cohen1962,Bardeen1962}. The first term in $I_S$, the $\sin(\phi)$ term, represents the standard Josephson effect (SJE). It arises as the tunneling connects the ground states of the two superconductors via virtual excitations. The $\cos(\phi)$ term is linked to the resistive component of the pair current, including a response in-phase with an alternating bias~\cite{Harris1974,Harris1975,Zorin1979,Pop2014}. 

We transcend the SJE, limited to smoothly-varying voltages, by exploring the microscopic Josephson response to sharply-varying voltage pulses and waves. This relatively unexplored regime~\cite{Perfetto2009,Stefanucci2010,Suoto2016} is invaluable for a precise account of high frequency applications. For instance, digital superconducting electronics, providing faster alternatives to their semiconducting counterparts~\cite{Likharev1991,Chen1999,Bunyk2001,Soloviev2017, Semenov2019,Ayala2021}, and superconducting quantum information processing employing electromagnetic pulses applied to Josephson tunnel junctions~\cite{Devoret2013,Makhlin2001,Clarke2008,Zazunov2003,Yamamoto2003,
Howe2022,Opremcak2021}. We discover that in Josephson tunnel junctions with topologically-trivial bound states (TTBSs) or Majorana zero modes (MZMs)~\cite{Kitaev2001,Kitaev2003,Nayak2008,Alicea2012,Lutchyn2010,Oreg2010,Lutchyn2018} subjected to sharply varying voltages, both the $\sin(\phi)$ and $\cos(\phi)$ terms are modified, and their interplay effectively yields a rich C$\phi$R. Specifically, for \emph{any} sub-gap state, we obtain a nonequilibrium fractional Josephson effect (NFJE) with the C$\phi$R $I_S(\phi)\sim\sin(\phi/2)$, oscillating at the \emph{fractional} Josephson frequency $\omega_J/2$. It initially dominates the SJE but eventually decays over the time-scale $\sim\hbar/\Gamma$, associated with the quasiparticle lifetime. For a topological Josephson tunnel junction hosting MZMs, the current has two \emph{independent} parts: (a) the usual parity-dependent MZM-induced fractional Josephson effect (MFJE)~\cite{Kitaev2001,Lutchyn2010}, typically obtained for a constant/smoothly-varying voltages~\cite{Wiedenmann2016,Deacon2017,Laroche2019,Dartiailh2021}, and, (b) the NFJE, which is universally found for any subgap states and sharply changing voltage. In contrast to the MFJE, which involves single quasiparticle tunneling, the NFJE originates from interfering \emph{two}-quasiparticle tunneling processes (see below) following sudden voltage changes. Consequently, it is independent of the ground-state fermion parity, establishing resilience against quasiparticle poisoning~\cite{Goldstein2011,Rainis2012,Budich2012,Karzig2021}. For experimental detection, we propose using a square wave bias to perpetually sustain the NFJE. Experimentally relevant time-scales and energies are provided below in the \textit{Discussion} section.

\begin{figure}
\begin{overpic}[height=4.8cm,width=0.63\linewidth]{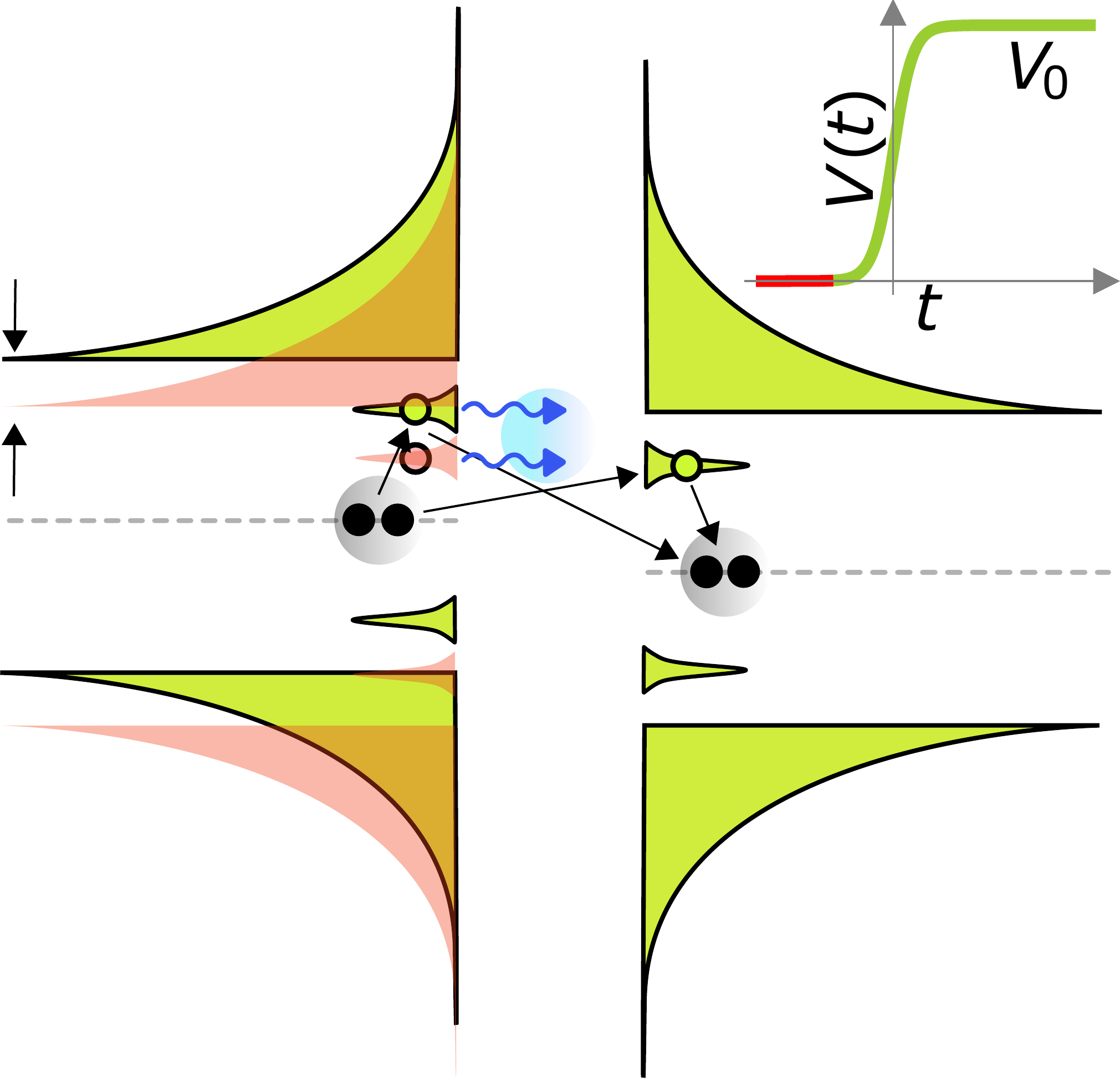}
 \put (4,70) {\large	$eV_0$}
 \put (42.5,66) {\large	$e^{i\omega_J't}$}
 \put (69.5,47.4) {\large  	$\omega_0$}
 \put (22,51.5) {\large		$\omega_0$}
\end{overpic} 
\caption{The dominant microscopic process for the Josephson pair current. A Cooper pair breaks up into two excitations (circles), here involving TTBSs at frequency $\omega_0$, combining after tunneling. The un-biased density of states (DOS) $(t\ll 0)$ is shown in red, while its biased counterpart $(t\gg 0)$ is shown in green. The interference (shaded blue) of the IQPs tunneling before and after the voltage step causes the $\omega_J'=\omega_J/2\pm 2\omega_0$ oscillations.}
\label{Fig1}
\end{figure}
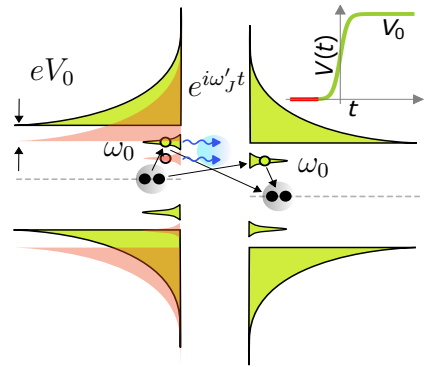

\textit{Phenomenology of the NFJE.--} We elucidate the key concepts underpinning the NFJE by considering the response to a voltage step rising to a height $V_0$ over time $\tau$. A microscopic calculation shows that the steady state pair current is dominated by Cooper pairs breaking into two Bogoliubov intermediate quasiparticles (IQPs), which tunnel across the junction and subsequently pair up (see Fig. \ref{Fig1}, and the Supplemental Material (SM)~\cite{sm}). The nonequilibrium pair current carries an imprint of this picture, with the dynamical response of IQPs dictating its temporal behaviour. Indeed, a sharply varying voltage excites a nonequilibrium distribution of IQPs, characterised by time and bias dependent phases, whose mutual interference~\cite{Gaury2014,Gaury2015,Wingreen1993,Wingreen1994} manifests as a nonequilibrium modulation of the amplitudes of the $\sin(\phi)$ and $\cos(\phi)$ terms at a fractional Josephson frequency $\omega_J'\approx\omega_J/2$.

In a heuristic description, at leading order in the tunneling amplitude, pair tunneling generates states with a single pair transferred across the junction in both directions. The pair current is given by the rate of change of their coefficients~\cite{sm}, with the resulting current oscillations determined by $J_p(t)= \int_{-\infty}^t dt' \exp[i(\phi(t)-\phi(t'))/2]\Im \iint_{-\infty}^\infty d\epsilon_L d\epsilon_R \nu(\epsilon_L)\nu(\epsilon_R)\exp[-i(\epsilon_R-\epsilon_L)(t-t')]  ( f(\epsilon_R) - f(\epsilon_L))$ where $\nu$ is the anomalous density of states~\cite{Bzdusek2013}, $\phi(t)/2=\int_{-\infty}^t dt' (e/\hbar)V(t')$, and $\Re J_P$ and $\Im J_P$ are the amplitudes of the $\sin(\phi)$ and $\cos(\phi)$ terms, respectively. This is the time-domain analogue of Fermi's golden rule. Its oscillatory response is determined by three predominant factors: (i) nonequilibrium excitations generated by the voltage step, (ii) interplay of $\cos(\phi)$ and $\sin(\phi)$ terms, and (iii) electronic retardation. We now provide a concise account of these factors. First, the IQPs at the junction ($x=0$) are characterised by the dynamic phase factor $e^{-i\chi(\epsilon,t)}$ with $\chi(\epsilon,t)=\epsilon t$. The voltage step alters it to $\chi'(t)=\chi(t)+\phi(t)/2$. Note that $\phi(t\gg\tau)/2\sim (\omega_J/2)t$ as $V(t\gg\tau)\to V_0$, while $\phi(t<0)=0$. Consequently, the IQPs tunneling at different times $t'$ acquire different phases $\phi$, commensurate with the time spent in the biased lead since the rise of the voltage step~\cite{sm}. The resulting mutual interference is captured by the factor $\exp[i(\phi(t)-\phi(t'))/2]$ in $J_p$, generating oscillations with frequency $\omega_J'\approx\omega_J/2$. Second, the $\cos(\phi)$ term typically requires a driving energy greater than the pair-breaking energy $E_{pb}$, which entails $V_0>E_{pb}$ for a constant bias $V(t)=V_0$~\cite{Barone1982}, or $\omega_d>E_{pb}$ for an alternating bias $V(t)=V_0\cos(\omega_d t)$~\cite{Harris1975}. However, in our dynamic scenario, the sharply varying voltage can supply $E_{pb}$. In particular, in the presence of low-energy TTBSs, apart from exhibiting the modulation described above, \emph{both} the $\sin(\phi)$ and $\cos(\phi)$ terms are of similar magnitudes. As such, they work concomitantly to suppress the underlying SJE and reinforce the NFJE. Third, the time-evolution of IQPs governed by their dynamic phases $\sim \exp(i(\epsilon_L+\epsilon_R)(t-t'))$ defines the microscopic retardation which, together with the first factor, determines $\omega_{J}'$. Consequently, the Josephson effect is enriched by exploring the electronic spectrum of the constituent superconductors. Additionally, the time-evolution is characterised by the inverse quasiparticle lifetime $\Gamma$, which dictates the longevity of the NFJE $\sim \hbar/\Gamma$. It naturally arises, e.g., from the inevitable relaxation to the high-energy quasiparticle continuum\mbox{~\cite{SanJose2012}}, external environments or dissipation\mbox{~\cite{Chamon2011,Budich2012,Rainis2012,Huang2020}}, etc. It also accounts for the coupling to the leads\mbox{~\cite{Budich2012,Pillet2010}} which, however, is relatively small in the tunnel limit.

Considering TTBSs at frequency $\omega_0$ in both leads serving as the IQP states, the bias-dependent and dynamic phases operate collectively to generate oscillations at frequency $\omega_{J,1/2}'=\omega_J/2\pm 2\omega_0$ in $J_p(t)$ (see Fig. S5 in~\cite{sm}), and the current. Notably, the dynamic phase vanishes for $\omega_0\to 0$, yielding the fractional Josephson frequency $\omega_J'=\omega_J/2$. For IQPs belonging to a band having a large band-width $\zeta\gg \omega_0$, their mutual interference washes out the NFJE oscillations after a short time $t\gtrsim \hbar/\zeta$, rendering them imperceptible. Therefore, we require TTBSs with \emph{sharp} spectral support in both leads, although not necessarily at the same frequency. The interference is dominated by IQPs tunneling presently at $t$ and those having tunneled in the recent past $t-\hbar/\Gamma\lesssim\tau<t$, particularly the ones which originated prior to the voltage step. Not only does this necessitate a voltage step varying faster than the quasiparticle lifetime, i.e., $\tau<\hbar/\Gamma$ (for smooth steps see~\cite{sm}), it also suppresses the NFJE for $t\gtrsim \hbar/\Gamma$ when the waves whose origin precede the voltage step have decayed. As such, $\Gamma$ cruicially determines the relevant time scales. 

\textit{Model.--} For a microscopic analysis, we follow the seminal work by Werthamer~\cite{Werthamer1966}, constituting a perturbative nonequilibrium formalism capable of handling arbitrary voltages. Considering a single-channel Josephson junction, with two s-wave superconducting leads coupled by the tunneling amplitude $\mathcal{T}$, we obtain the current
\begin{equation}
\begin{split}I(t)=  \frac{e\mathcal{T}^2}{\hbar}\int\displaylimits_{-\infty}^t dt' \bigg[  &\sin\bigg(\frac{\phi(t)+\phi(t')}{2}\bigg)K_S(t-t')\\
-&\sin\bigg(\frac{\phi(t)-\phi(t')}{2}\bigg)K_N(t-t')\bigg].
\end{split}
\label{iwerttime2}
\end{equation}

The first and second lines represent the pair and quasiparticle currents, respectively. $K_{N/S}$ are the response kernels describing the retardation, given by $K_{N/S}(t)=\Im \int d\epsilon_L d\epsilon_R e^{-i(\epsilon_L-\epsilon_R)t}A^{N/S}_L(\epsilon_L)A^{N/S}_R(\epsilon_R)[f(\epsilon_R)-f(\epsilon_L)]$, where $f$ is the Fermi function, and $A^N_j=-[\textbf{Im}G^r_j]_{1,1}$ and $A^S_j=-[\textbf{Im}G^r_j]_{1,2}$ are normal and anomalous components of the surface spectral function of the lead $j=L(\text{left})/R(\text{right})$. They contain the dynamic phases representing the time-evolution of IQPs, weighted by the corresponding spectral functions. Eq.\eqref{iwerttime2}, in turn, represents the interference between the tunneled state at $t'$ and its time-evolved version at the present time $t$, considering both the retardation from $K_{N/S}$ and the bias-dependent phases, thereby accounting for the mutual interference of excited IQPs~\cite{sm}. In the presence of TTBSs, noting that $A^S_j(\omega)$ is odd in $\omega$ for s-wave superconductors~\cite{Bzdusek2013}, $A^{N/S}_j(\omega)=A^{N/S}_{j,\text{band}}(\omega)+h_{j,n/s}(4\Delta/\zeta)[\delta(\omega-\omega_0)\pm\delta(\omega+\omega_0)]$, where $\omega_0$ and $h_{j,n/s}$ are frequency and spectral weight of the TTBS, respectively, $\Delta_L=\Delta_R=\Delta\ll\zeta$ is the superconducting gap, and $A^{N/S}_{j,\text{band}}$ corresponds to band states. The TTBS peak is regularised by $0<\Gamma\ll \Delta$. We relegate the expressions for $K_{N/S}$ to the SM~\cite{sm} for brevity. For the bands, $K_{N/S}$ oscillates at frequency, $\omega=2\Delta/\hbar$~\cite{Harris1975,Choi2022}, corresponding to the singular band-edge spectral density. Additionally, being associated with the energy scale $\Delta$, it follows from the uncertainty principle that it is significant only for a typically short time $\sim \hbar/\Delta$. However, when both leads have TTBSs with $\Gamma\ll\Delta$, $K_{N/S}$ is significant for a longer time $\sim\hbar/\Gamma\gg \hbar/\Delta$. Moreover, corresponding to the dynamical evolution of the TTBS, it contains oscillations associated with its characteristic frequency $\omega_0$.

In the case of topological leads with MZMs, the four-fold ground state degeneracy associated with the parities of each disconnected lead entails a careful choice of the the initial junction ground state for a convergent perturbative result\mbox{~\cite{Brouder2009,sm}}. This creates parity-dependent correlations in the junction ground state, leading to the parity-dependent MFJE\mbox{~\cite{Kitaev2001,Lutchyn2010}} at order $\mathcal{T}$, $I^{(0)}=(e\mathcal{T}/\hbar)\mathpzc{p}_I\sin(\phi(t)/2)\psi_{L,1}\psi_{R,N}/2$, where $\psi_{L,1}(\psi_{R,N})$ represents the left(right)-localised MZM wavefunctions evaluated at the left(right) end of the wire, and $\mathpzc{p}_I$ is the fermion-parity defined by the two MZMs located near the junction. At order $\mathcal{T}^2$, we again obtain Eq. \mbox{\eqref{iwerttime2}}. Remarkably, unlike $I^{(0)}$, Eq. \eqref{iwerttime2} is independent of the choice of the ground state, and hence, the parity. The reason is that the NFJE is governed by the mutual interference in double IQP tunneling at different times (see Fig. \mbox{\ref{Fig1}} and SM\mbox{~\cite{sm}}). While each single IQP process depends on parity, like the MFJE, double IQP tunneling is independent of the parity of ground states. As such, the current remains unaffected by parity-flipping processes. Furthermore, for a statistically mixed ensemble of ground states with different parities, which may be established by incoherent transitions to the environment, the NFJE current survives while $I^{(0)}$ averages out to zero. We explore these points further in the SM~\cite{sm}.

\begin{figure}
\begin{overpic}[clip,width=0.99\columnwidth]{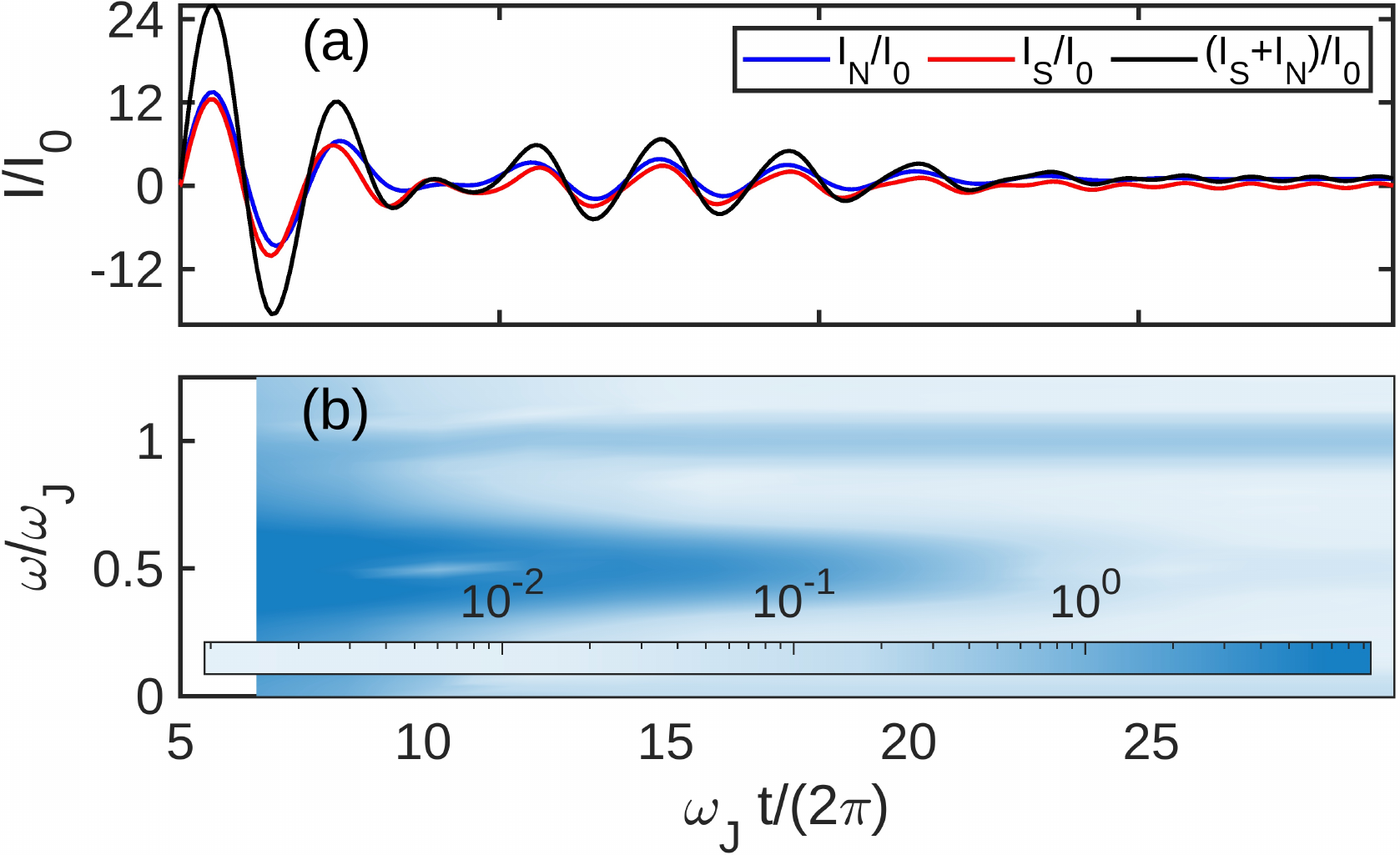}
 \put(72.9,37.8){\tikz \draw[line width=0.6mm,dashed,gray](60,11.2)--(60,9.9);}
 \put(72.9,15.4){\tikz \draw[line width=0.6mm,dashed,gray](60,8.41)--(60,9.94);}
 \put(73,40.5){\tikz \draw [-{Stealth[scale=2,gray]}] (0,0) -- (1,0);} 
 \put(88.5,40.5) {\large SJE}
 \put(62,40.5){\tikz \draw [-{Stealth[scale=2,gray]}] (0.0,0) -- (-1,0);} 
 \put(45,40.5) {\large NFJE}
\end{overpic}
\caption{(a) Normalised current response to a step bias $V(t)=V_0\Theta(t)$, with $eV_0=(\pi/2)\Delta$, $\zeta=150\Delta$, $\Gamma=0.1\Delta$, and $I_0=(e\mathcal{T}^2/\hbar)(2\Delta^2/\hbar^2\zeta^2)(\hbar/\Delta)=\pi\Delta/(8eR_N)$. We consider TTBSs in both leads at $\omega_0=0.1\Delta$ with $h_n=h_s=1$~\cite{bandneg}. Strong $\omega_J/2-$oscillations dominate initially, eventually decaying to SJE oscillations. (b) Short-time Fourier transform of the current in (a), showing the frequency components versus time, revealing the NFJE$(\omega_J/2)\to$ SJE$(\omega_J)$ transition at time $\sim\hbar/\Gamma$ (dashed line). The \textit{Discussion} section provides an estimate for it, which determines the remaining parameters}
\label{Fig2}
\end{figure}
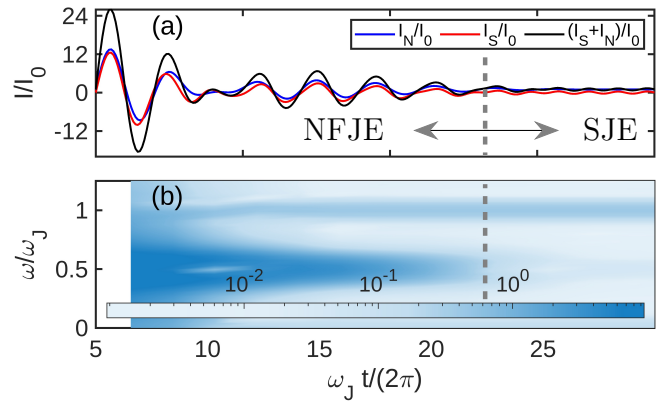

\textit{Heaviside step pulse.--} Here we elaborate on the response to a Heaviside step voltage, $V(t)=V_0\Theta(t)$, focusing on TTBSs and MZMs. For $t>0$, the pair current is obtained from Eq. \eqref{iwerttime2} as
\begin{equation}
I_S(t)=\frac{e\mathcal{T}^2}{\hbar}\bigg[\sin\bigg(\frac{2eV_0t}{\hbar}\bigg)\Re J_p(t)-\cos\bigg(\frac{2eV_0t}{\hbar}\bigg)\Im J_p(t)\bigg],\label{iwerttimes}
\end{equation} 
where
\begin{equation}
J_p(t)=\int_{-\infty}^t dt'e^{i\frac{(\phi(t)-\phi(t'))}{2}}K_S(t-t').
\end{equation}
$\Re J_p$ and $\Im J_p$ define the amplitude of the $\cos(\phi)$ and $\sin(\phi)$ terms, respectively. The oscillations in $J_p$ arise from a combination of the bias-dependent excitation and retardation of the IQPs. An analytical approximation~\cite{sm} valid for $eV_0\gg\omega_0\gg\Gamma$, shows that $I_S(t)$ is initially dominated by NFJE oscillating at $\omega_{J,1/2}'=\omega_J/2\pm 2\omega_0$ with an amplitude $\sim (e\mathcal{T}^2/\hbar)(2\Delta^2/\hbar^2\zeta^2)(\hbar/\Gamma)=I_0(\Delta/\Gamma)$, which transitions into the SJE over a time $\sim\hbar/\Gamma$. Here, $I_0=(e\mathcal{T}^2/\hbar)(2\Delta^2/\hbar^2\zeta^2)(\hbar/\Delta)=\pi\Delta/(8eR_N)$ is the SJE critical current arising from the band states\mbox{~\cite{sm}}, with $R_N$ being the normal state resistance. These predictions follow from the temporal behaviour of $K_{S}$, as described earlier. Importantly, the NFJE is parametrically larger than the SJE by the factor $\Delta/\Gamma$. For a given voltage $V_0$, since $K_{N/S}$ contains oscillations associated with $\omega_0$, the NFJE response improves with decreasing $\omega_0$~\cite{sm}. Hence, a high voltage $eV_0>>\hbar\omega_0$ alongside $eV_0>>\Gamma$ yields a pronounced NFJE at frequency $\omega_J/2$. These features are shown in Fig. \ref{Fig2} for a single TTBS in each lead~\cite{bandneg}. Note that TTBSs with $\omega_0\lesssim\Gamma$, and MZMs, naturally satisfy this criterion. We remark that the $\sin(\phi)-$term, on its own, is insufficient to generate NFJE, instead creating beatings containing both $\omega_J$ and $\omega_J/2-$oscillations. A coherent interference between the $\cos(\phi)$ and $\sin(\phi)$ channels is required for NFJE composed solely of $\omega_J/2-$oscillations.

\begin{figure}
\includegraphics[height=4.7cm,width=0.99\columnwidth]{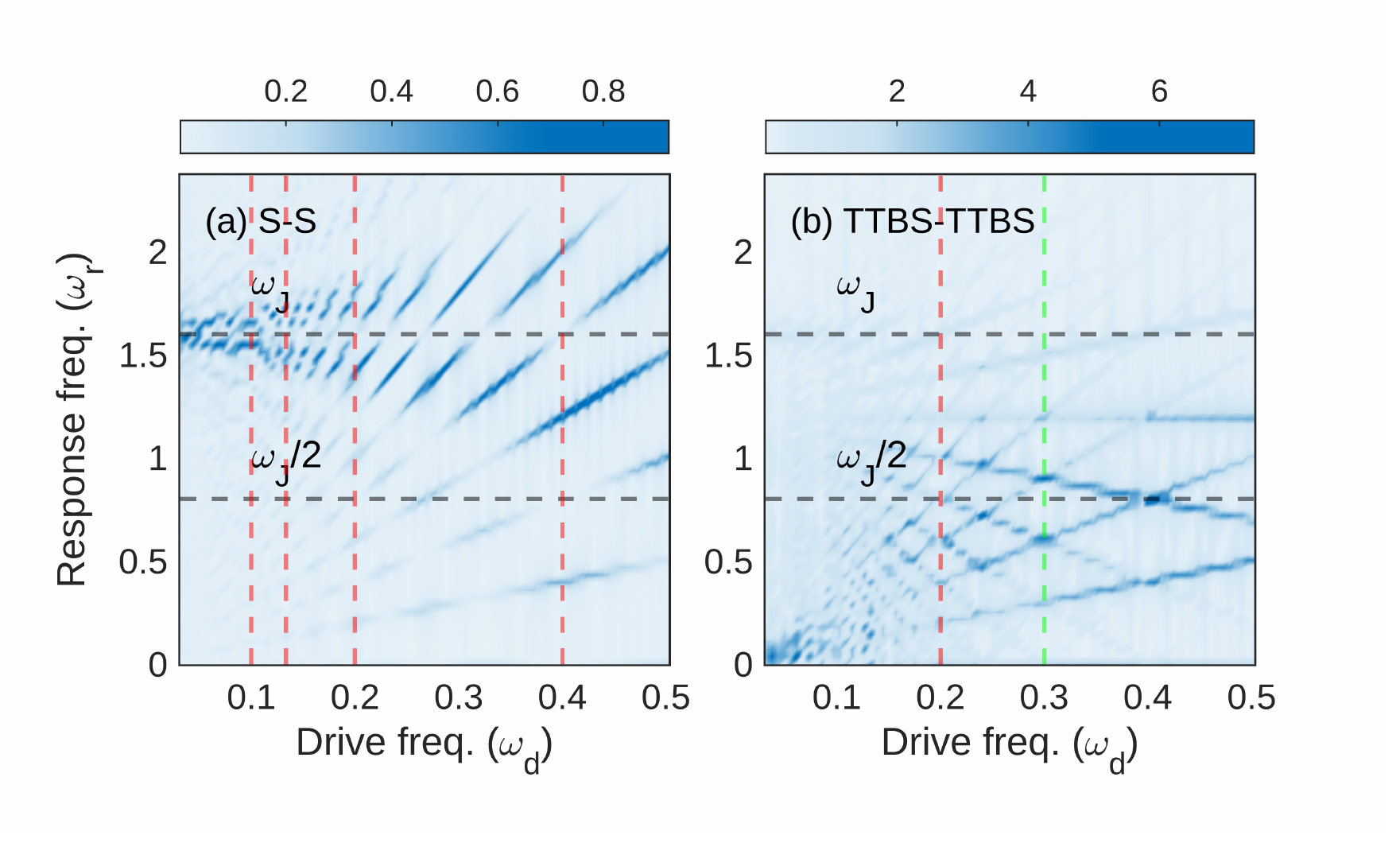}
\caption{Normalised current $(I/I_0)$ response to a square wave bias with $eV_0=0.8\Delta$ and duty-cycle $D=0.75$, resolved in Fourier domain. In (a) we assume BCS superconducting leads, and in (b) TTBSs at $\omega_0=0.1\Delta$ in both leads~\cite{bandneg} with $\Gamma=0.1\Delta$. For $\omega_d\gtrsim \Gamma/\hbar$, the NFJE oscillations dominate each pulse cycle in the presence of TTBSs, whereas for $\omega_d\lesssim \Gamma/\hbar$, they decay to reveal the static normal current through the TTBS. The stripes arise due to the non-linear response to the periodic voltage drive, creating resonances. Some of the prominent ones are marked in green(red), corresponding to positive bias(both) section(s) of each square pulse being commensurate with the NFJE oscillation period.}
\label{Fig3}
\end{figure}
For a topological Josephson junction hosting MZMs, the current admits a compact analytical approximation in the wideband limit ($\zeta\gg\Delta,eV_0$) for $\Gamma\ll eV_0$~\cite{sm},
\begin{equation}
I_S(t)=I_0\left[\left(\frac{\Delta}{\Gamma}\right) \frac{1/2}{1+(\frac{\Gamma t}{\hbar})^2}\sin\left(\frac{\omega_J}{2}t\right)+\mathcal{O}\left(\frac{\Delta}{\Gamma}\right)^0\right].\label{nfjemzm}
\end{equation} 
The sub-leading term, which is parametrically smaller than the NFJE current by the factor $\Gamma/\Delta$, is the band-state SJE current.

\textit{Square wave bias.--}  For experimental detection, the NFJE can be sustained by periodically driving it with a square wave bias varying between zero and a positive voltage, as shown in Fig. \ref{Fig3}. For this, we require a drive frequency $\omega_d\gtrsim \Gamma/\hbar$, such that new NFJE pulses are generated before the preceding ones decay. Additionally, the commensurability of the NFJE oscillations with the voltage drive, along with the non-linearities of the system, generate several resonances, letting us choose $\omega_d$ and the fraction $D$ of the square pulse having positive voltage (duty cycle) to strengthen the NFJE signal. Specifically, the strongest resonances are obtained when the durations of both sections of each square pulse are commensurate with the NFJE oscillation period $2\pi \hbar/eV$. This translates to the requirement $(2\pi D/\omega_d)/(2\pi\hbar/eV),(2\pi (1-D)/\omega_d)/(2\pi\hbar/eV)\in \mathbb{Z}$, which ensures that the pair current presents a repeatable form in each pulse, and is in-phase with the normal current. Further examples, including the case of MZMs, are presented in the SM~\cite{sm}.

\begin{figure}
\includegraphics[clip,width=0.99\columnwidth]{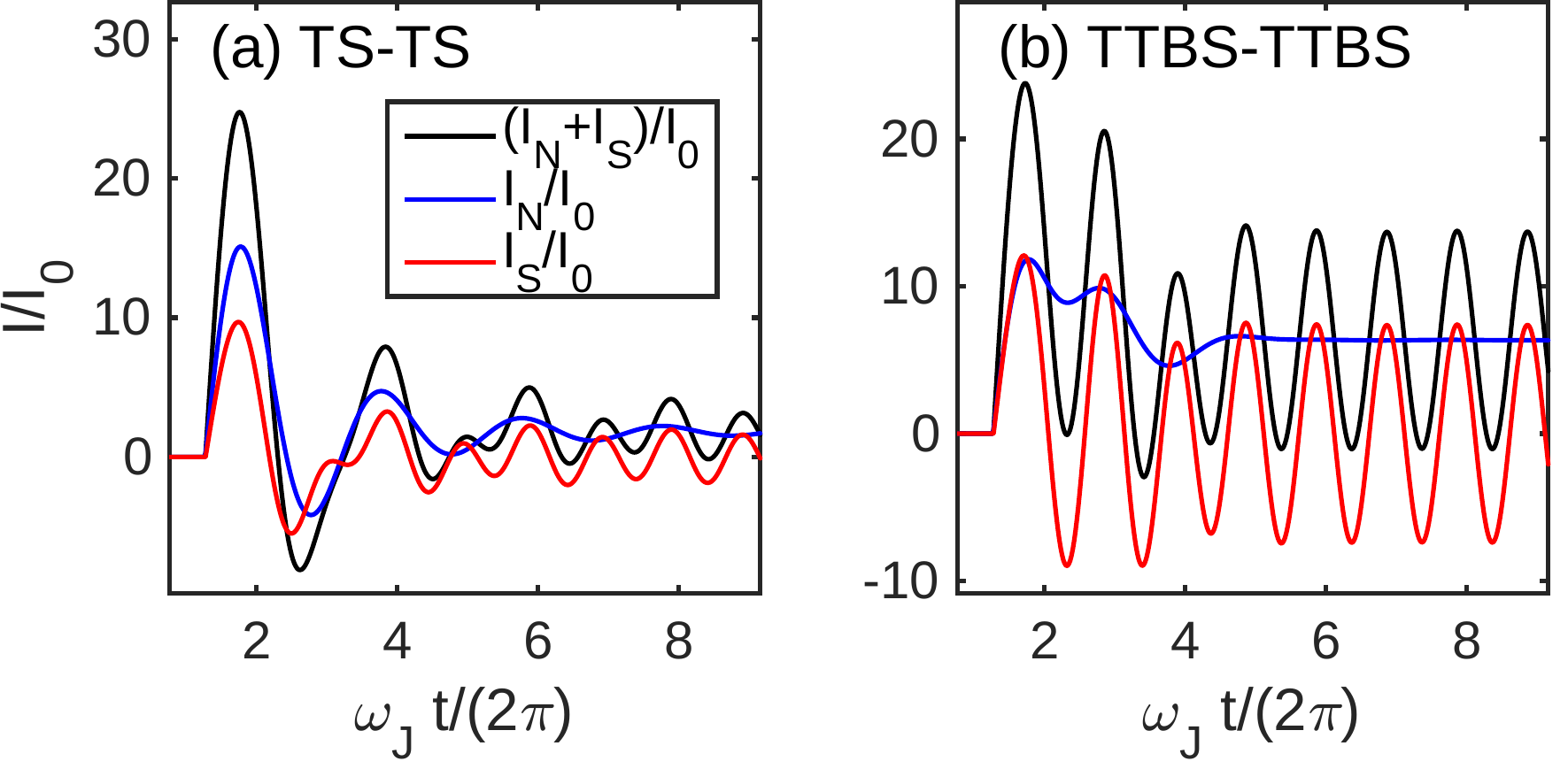}
\caption{Normalised current response to a small-amplitude step bias $V(t)=V_0\Theta(t)$, with $eV_0=0.4\Delta$, $\zeta=150\Delta$, and $\Gamma=0.1\Delta$. In (a), we consider topologically superconducting (TS) leads based on Kitaev chains, and in (b) we consider TTBSs in both leads at $\omega_0=0.1\Delta=\Gamma$ with $h_n=h_s=1$~\cite{bandneg} ($eV_0=4\hbar\omega_0$). Unlike the TS junction in (a), there is no noticeable NFJE for the TTBSs in (b).}
\label{Fig4}
\end{figure}

\textit{Discussion.--} 
Interestingly, unlike conventional transport techniques whose resolution is limited by the broadening $\Gamma$, NFJE can distinguish MZMs from low-energy TTBSs even with a large broadening $\Gamma\sim\hbar\omega_0\ll\Delta$~\cite{Chen2019,Contamin2022}. This is because the NFJE frequency, which accounts for both the interference $(\sim\omega_J/2)$ and time-evolution $(\sim\omega_0)$ of IQPs, embodies the energy scales as $\omega'_{J,\text{MZM}}=\omega_J/2$ for MZMs and $\omega'_{J,\text{TTBS}}=\omega_J/2-2\omega_0$ for the TTBSs (see Fig. S5 in\mbox{~\cite{sm}}). These energy scales can be significantly different at low voltages comparable to $\omega_0$. Since the NFJE lifetime is $\sim\hbar/\Gamma$, we impose $2\pi/\omega_{J,\text{TTBS}}\gtrsim \hbar/\Gamma$ to render the NFJE from TTBSs inconspicuous as it decays rapidly within a single oscillation period, whereas $2\pi/\omega_{J,\text{MZM}}\lesssim \hbar/\Gamma$, leading to noticeable NFJE oscillations from MZMs. Practically, for $\hbar\omega_0\approx\Gamma$, we find $eV_0=\alpha\hbar\omega_0$ with $\alpha\approx 3-5$, as shown in Fig. \mbox{\ref{Fig4}}. Following the previous section, we can employ a square-wave voltage with this amplitude to experimentally achieve this distinction\mbox{~\cite{sm}}. 

For the experimental realisation of the NFJE, we require low-energy TTBSs in both leads with sharp spectral support, which may arise from disorder~\cite{Pan2020,Pan2021a,Pan2021}, inhomogeneities in the chemical potential or the superconducting gap~\cite{Kells2012,Liu2017,Reeg2018,Moore2018,Moore2018a}, as Andreev bound states (ABS) in SNS junctions, Shiba states in presence of magnetic adatoms~\cite{Yu1965,Shiba1968,Rusinov1969}, or MZMs. Primarily, the voltage rise time $\tau$ must satisfy $\tau<\hbar/\Gamma$. Experiments on Shiba states~\cite{Ruby2015,Huang2020,Thupakula2022} have revealed $\hbar/\Gamma\sim 0.4-40$ns, with the larger values typically associated with lower temperatures. A similar value, $\sim 1$ns, was reported for topological gapless ABS~\cite{Deacon2017}. Other studies~\cite{Rainis2012,Janvier2015,Hays2018,Zellekens2022,elfeky2023,Higginbotham2015} have estimated relaxation times $1\mu$s$-10$ms, from quasiparticle poisoning. Considering a conservative estimate of $\hbar/\Gamma\sim 1$ns, the NFJE is accessible with current experiments. In fact, voltage pulses with $\tau\sim 50$ps have already been demonstrated~\cite{Dubois2013,Duboisthesis,Gaury2015}. For a pronounced NFJE, we require $\Delta\gg eV_0\gg\hbar\omega_0,\Gamma$, where the first inequality limits hot quasiparticles excited by the bias which may degrade the lifetime. Using, for instance, HgTe-based topological Josephson junctions~\cite{Deacon2017} or NbTiN/InSb hybrid nanowire devices~\cite{Chen2019} as a guide, we typically have $\Delta\sim 100-300\mu$eV and $\Gamma/\Delta\sim 0.01-0.1$, leaving substantial room for suitable $eV_0$ and $\omega_0$. In particular, following the preceding discussion, considering TTBSs at $\hbar\omega_0=\Gamma$, we require a square wave with $eV_0\sim 5-50 \mu$eV$\ll\Delta$, and drive frequency $\omega_d/(2\pi)\sim(0.5-5)$GHz to distinguish them from MZMs.

\begin{acknowledgements}
This work was supported by the W\"urzburg-Dresden Cluster of Excellence ct.qmat, EXC2147, project-id 390858490, and the DFG (SFB 1170). We thank the Bavarian Ministry of Economic Affairs, Regional Development and Energy for financial support within the High-Tech Agenda Project ``Bausteine f\"ur das
Quanten Computing auf Basis topologischer Materialen."
\end{acknowledgements}

\let\oldaddcontentsline\addcontentsline
\renewcommand{\addcontentsline}[3]{}

\let\addcontentsline\oldaddcontentsline

\pagebreak
\onecolumngrid

\makeatletter

\renewcommand{\thesection}{S\arabic{section}}
\renewcommand{\theequation}{S\arabic{equation}}
\renewcommand{\thefigure}{S\arabic{figure}}
\renewcommand{\bibnumfmt}[1]{[S#1]}
\renewcommand{\citenumfont}[1]{S#1}

\newpage
\begin{large}
\begin{center}
\textbf{Supplemental Material: Nonequilibrium Fractional Josephson Effect} 
\end{center}  
\end{large}

{\justifying
In this Supplemental Material, we present and discuss more details for (S1) the Hamiltonian employed in this work; (S2) the derivation of the tunneling current for the s-wave and p-wave superconductors, including a phenomenological description of the current, the theoretical procedure to handle the topological degeneracy in the latter, and the parity dependence; (S3) low-energy Andreev bound states (ABS) in inhomogeneous s-wave superconducting nanowires; (S4) additional data for the Josephson response to Heaviside step bias (and its smoother version), including the cases of ABS and Majorana modes; (S5) additional data for the response to-square wave bias; (S6) experimental distinction of MZMs and TTBSs using a square-wave bias.
}
\setcounter{page}{1}
\setcounter{secnumdepth}{3}
\setcounter{section}{0}
\setcounter{equation}{0}
\setcounter{figure}{0}
\setcounter{table}{0}

\tableofcontents
\section{Hamiltonian}
In this section, we specify the Hamiltonians used in this work. For both the conventional BCS superconductor and topological Kitaev chain, the factors multiplying $\Delta$ and $\zeta$ are chosen to ensure that the spectral gap and bandwidth are $2\Delta$ and $2\zeta$, respectively.
\begin{itemize}
\item s-wave superconductor 
\begin{align}
H_{L/R}=&\sum_j\sum_\sigma \bigg[\frac{1}{2}\zeta\Big(-c_{j+1,\sigma}^\dagger c_{j,\sigma}-c_{j,\sigma}^\dagger c_{j+1,\sigma}\Big)+\frac{1}{2}\Big(\Delta^*[-i\sigma_2]_{\sigma\sigma'} c_{j,\sigma}c_{j,\sigma'}+\Delta[i\sigma_2]_{\sigma\sigma'}c^\dagger_{j,\sigma}c^\dagger_{j,\sigma'}\Big)-\mu c_{j,\sigma}^\dagger c_{j,\sigma}\bigg]\label{Hsw}.
\end{align}
with excitation spectrum $\omega_k=\sqrt{(\zeta\cos(ka)+\mu)^2+\Delta^2}$.
\item Kitaev chain topological superconductor (TS)
\begin{align}
H_{L/R}=&\sum_j \bigg[\frac{1}{2}\zeta\Big(-c_{j+1}^\dagger c_{j}-c_{j}^\dagger c_{j+1}\Big)+\frac{1}{2}\Big(\Delta c_{j}c_{j+1}+\Delta^*c^\dagger_{j+1}c^\dagger_{j}\Big)-\mu c_{j}^\dagger c_{j}\label{Hpw}
\end{align}
with excitation spectrum $\omega_k=\sqrt{(\zeta\cos(ka)+\mu)^2+(\Delta\sin(ka))^2}$.
\end{itemize}

The p-wave superconductor with a generic superconducting phase belongs to class D. In this case, we have the charge-conjugation operator $\mathcal{P}=\tau_x\mathcal{K}$, such that $\mathcal{P}H_{\text{BdG}}\mathcal{P}^{-1}=-H_{\text{BdG}}$. The Hamiltonian is diagonalised by,
\begin{align}
\begin{bmatrix}
c_1 \\ c_1^\dagger \\ c_2 \\ c_2^\dagger \\ \vdots
\end{bmatrix} = & \underbrace{\begin{bmatrix}
\psi_{L,1} & i\psi_{R,1} & v_1^2 & u_1^2 & \ldots\\
\psi_{L,1} & -i\psi_{R,1} & {u_1^2}^* & {v_1^2}^* & \ldots\\
\psi_{L,2} & i\psi_{R,2} & v_2^2 & u_2^2& \ldots\\
\psi_{L,2} & -i\psi_{R,2} & {u_2^2}^* & {v_2^2}^* & \ldots\\
\vdots& \vdots& \vdots& \vdots& \ddots
\end{bmatrix}}_{S}
\begin{bmatrix}
\gamma_1 \\ \gamma_2 \\ d_2 \\ d_2^\dagger \\ \vdots
\end{bmatrix},
\end{align}
where $\psi_{L,j}$ and $\psi_{R,j}$ are real. The first two columns in $S$, correspond to the Majorana zero modes (MZM), which are their own particle-hole partners. This is in contrast to the case of non-Majorana quasiparticles where the $2j^{\text{th}}$ and $(2j-1)^{\text{th}}$ columns are particle-hole partners of each other for all $j>1$. We denote that $\psi_{L}$ and the corresponding operator $\gamma_1$ represent the left-localised MZM, whereas $\psi_{R}$ and $\gamma_2$ denote the right-localised MZM. In sufficiently long chains consisting of $N$ sites, we have $\psi_{L,N}\to0$ and $\psi_{R,1}\to 0$.

In order to define the ground state, we require the fermionic operators $d_1=(\gamma_1-i\gamma_2)/\sqrt{2}$ and $d_1^\dagger=(\gamma_1+i\gamma_2)/\sqrt{2}$ which satisfy the fermionic anti-commutation relations, with the even-parity ground state satisfying $d_j|g_e\rangle=0$ $\forall j=1\ldots N$. The second/odd-parity ground state is given by $|g_o\rangle=d_1^\dagger|g_e\rangle$. We thus have in this new basis, 
\begin{align}
\begin{bmatrix}
c_1 \\ c_1^\dagger \\ c_2 \\ c_2^\dagger \\ \vdots
\end{bmatrix} = & \underbrace{\begin{bmatrix}
\frac{\psi_{L,1}+\psi_{R,1}}{\sqrt{2}} & \frac{\psi_{L,1}-\psi_{R,1}}{\sqrt{2}} & v_1^2 & u_1^2 & \ldots\\
\frac{\psi_{L,1}-\psi_{R,1}}{\sqrt{2}} & \frac{\psi_{L,1}+\psi_{R,1}}{\sqrt{2}} & {u_1^2}^* & {v_1^2}^* & \ldots\\
\frac{\psi_{L,2}+\psi_{R,2}}{\sqrt{2}} & \frac{\psi_{L,2}-\psi_{R,2}}{\sqrt{2}} & v_2^2 & u_2^2& \ldots\\
\frac{\psi_{L,2}-\psi_{R,2}}{\sqrt{2}} & \frac{\psi_{L,2}+\psi_{R,2}}{\sqrt{2}} & {u_2^2}^* & {v_2^2}^* & \ldots\\
\vdots& \vdots& \vdots& \vdots& \ddots
\end{bmatrix}}_{S}
\begin{bmatrix}
d_1 \\ d_1^\dagger \\ d_2 \\ d_2^\dagger \\ \vdots
\end{bmatrix}.\label{Skcrot}
\end{align}

\subsection{Spectral function}
The spectral functions can be obtained by calculating the surface Green's function recursively using the Dyson equation~\cite{sPeng2017}. Since we work in the wide-band limit $(\zeta\gg\Delta)$ for simplicity, we can obtain compact analytical expressions instead~\cite{sZazunov2016}, $A^N=-(1/2)\Im \mathbf{Tr}[G\tau_0]$ and $A^S=-(1/2)\Im \mathbf{Tr}[G\tau_x]$, where $\tau_j$ are the Pauli matrices.
\begin{itemize}
\item s-wave superconductor: BCS superconductor
\begin{align}
A^N_{L/R}\tau_0+A^S_{L/R}\tau_x=&\frac{4\Theta(|\omega|-\Delta)}{\zeta\sqrt{\omega^2-\Delta^2}}\big(|\omega|\tau_0+\Delta\text{sign}(\omega)\tau_x\big).
\end{align}
\item p-wave superconductor (TS): Kitaev chain
\begin{align}
A^N_{L/R}\tau_0+A^S_{L/R}\tau_x=&\bigg(\frac{4}{\zeta}\Delta\pi\delta(\omega)[\tau_0\pm\tau_x]+\frac{4}{\zeta}\frac{\sqrt{\omega^2-\Delta^2}}{|\omega|}\tau_0\Theta(|\omega|-\Delta)\bigg),
\end{align}
Note that the MZM peak in the anomalous spectral function $B$ has opposite signs for the two leads. Also, the fact that the anomalous spectral function contains only MZM contribution $\sim\delta(\omega)$ is true only in the wideband limit.
\end{itemize}

\newpage
\section{Tunneling Current}
\subsection{s-wave (BCS) superconductor}
In this section, we follow Werthamer~\cite{sWerthamer1966} in deriving the current. On gauging out the applied bias potential (positive voltage $V(t)$ applied to the left lead) by making the unitary transformation $c\to ce^{i\phi(t)/2} $, with $d\phi(T)/dt=2eV(t)/\hbar$, all the bias dependence is shifted into the tunnel amplitudes. This yields the tunneling Hamiltonian,
\begin{align}
H_T=& \sum_\sigma -\mathcal{T}\big(e^{-i\phi(t)/2}c_{L,1\sigma}^\dagger c_{R,1\sigma}+e^{i\phi(t)/2}\hat{c}_{R,1\sigma}^\dagger c_{L,1\sigma}\big),\label{HT}
\end{align}
where $\mathcal{T}$ is the tunnel coupling, and $\phi(t)=\phi_L(t)-\phi_R(t)$ is the superconducting phase difference between the left and the right leads. We have indexed the sites such that site 1 from each lead is nearest to the junction. Hence, the current operator is obtained as,
\begin{align}
\hat{I}=&\sum_\sigma\frac{ie(-\mathcal{T})}{\hbar}\big(e^{-i\phi(t)/2}c^\dagger_{L,1,\sigma}c_{R,1,\sigma}-e^{i\phi(t)/2}c^\dagger_{R,1,\sigma}c_{L,1,\sigma}\big),
\end{align}
where in the pre-factor $e=|e|$ is the unsigned electronic charge. Within the Keldysh formalism, this is evaluated as,
\begin{align}
I=&\sum_\sigma\frac{e(-\mathcal{T})}{\hbar}\Re\Big[e^{-i\phi(t)/2}G_{R,1,L,1,\sigma}^{+-}(t,t)\Big].
\end{align}
For a pedagogical guide, see, for instance, Refs.~\cite{sCuevasbook2017,sJauhobook2008,sMeir1992}. Expanding up to order $\mathcal{T}^2$,
\begin{align}
I=&-\frac{e(-\mathcal{T})^2}{\hbar}\sum_\sigma\int_{-\infty}^\infty dt_1\Theta(t-t_1)\Re \bigg[e^{-i\frac{(\phi(t)-\phi(t_1))}{2}}\Big(g_{R,1,R,1,\sigma}^{-+}(t,t_1)g_{L,1,L,1,\sigma}^{+-}(t_1,t) - g_{R,1,R,1,\sigma}^{+-}(t,t_1)g_{L,1,L,1,\sigma}^{-+}(t_1,t)\Big)\nonumber\\
&\hspace*{41.5mm}- e^{-i\frac{(\phi(t)+\phi(t_1))}{2}}\Big( \tilde{f}_{R,1,R,1,\sigma\sigma'}^{-+}(t,t_1)f_{L,1,L,1,\sigma'\sigma}^{+-}(t_1,t)- \tilde{f}_{R,1,R,1,\sigma\sigma'}^{+-}(t,t_1)f_{L,1,L,1,\sigma'\sigma}^{-+}(t_1,t)\Big)\bigg],\label{Isstemp}
\end{align}
where $\tilde{f}_{a,b,\sigma,\sigma'}^{\alpha\beta}(t,t_1)=-iT_K\langle c_{a,\sigma}(t^\alpha) c_{b,\sigma'}(t_1^\beta) \rangle$ and $f_{a,b,\sigma,\sigma'}^{\alpha\beta}(t,t_1)=-iT_K\langle {c^\dagger_{a,\sigma}}(t^\alpha) {c^\dagger_{b,\sigma'}}(t_1^\beta) \rangle$, with $T_K$ being the time-ordering rule on the Keldysh contour. This expression is easier to evaluate in the Fourier domain. Defining $e^{-i\phi(t)/2}=\int (d\omega/2\pi)W(\omega)e^{-i\omega t}$, we obtain,
\begin{align}
I=&-\frac{e\mathcal{T}^2}{\hbar}\Re \iint_{-\infty}^{\infty}\frac{d\omega}{2\pi}\frac{d\omega'}{2\pi}\bigg(W(\omega)W^*(\omega')e^{-i(\omega-\omega')t} Q^N(-\omega')-W(\omega)W(\omega')e^{-i(\omega+\omega')t}Q^S(\omega')\bigg) \label{ineqssw}
\end{align}
where,
\begin{subequations}
\begin{align}
Q^{N/S}(\omega')=&\iint_{-\infty}^\infty\frac{d\omega_1}{2\pi}\frac{d\omega_2}{2\pi}\frac{i}{\omega_2-\omega_1+\omega'+i\eta} A^{N/S}_L(\omega_1)A^{N/S}_R(\omega_2)[f(\omega_2)-f(\omega_1)],\label{qsw}
\end{align}
\end{subequations}
On Fourier transforming, 
\begin{align}
I(t)=&-\frac{e\mathcal{T}^2}{\hbar}\Re \int_{-\infty}^\infty d\tau \big(e^{-i(\phi(t)-\phi(t-\tau))/2}Q^N(\tau)-e^{-i(\phi(t)+\phi(t-\tau))/2}Q^S(\tau)\big),\label{IQt}
\end{align}
where $Q^{N/S}(t)=\Theta(t)\mathcal{K}_{N/S}(t)$, with the kernels,
\begin{align}
\mathcal{K}_{N/S}(t)\coloneqq iK_{N/S}(t)=&\int_{-\infty}^\infty\frac{d\omega'}{2\pi}e^{-i\omega't}\int_{-\infty}^\infty\frac{d\Omega}{2\pi}A^{N/S}_L\Big(\Omega+\frac{\omega'}{2}\Big)A^{N/S}_R\Big(\Omega-\frac{\omega'}{2}\Big)\bigg[f\Big(\Omega-\frac{\omega'}{2}\Big)-f\Big(\Omega+\frac{\omega'}{2}\Big)\bigg]\label{knksgen},\\
=&\iint_{-\infty}^\infty\frac{d\epsilon_L}{2\pi}\frac{d\epsilon_R}{2\pi}e^{-i(\epsilon_L-\epsilon_R)t}A^{N/S}_L(\epsilon_L)A^{N/S}_R(\epsilon_R)\big[f(\epsilon_R)-f(\epsilon_L)\big].\label{knksgen1}
\end{align}
Finally, the time-domain current is obtained as,
\begin{align}
I(t)=&-\frac{e\mathcal{T}^2}{\hbar}\Re \int_{0}^\infty d\tau\bigg(e^{-i\frac{(\phi(t)-\phi(t-\tau))}{2}}iK_N(\tau)-e^{-i\frac{(\phi(t)+\phi(t-\tau))}{2}}iK_S(\tau)\bigg),\label{ineqssf0}\\
=&\frac{e\mathcal{T}^2}{\hbar} \int_{-\infty}^t dt_1 \bigg[  -\sin\bigg(\frac{\phi(t)-\phi(t_1)}{2}\bigg)K_N(t-t_1)+\sin\bigg(\frac{\phi(t)+\phi(t_1)}{2}\bigg)K_S(t-t_1)\bigg]. \label{ineqssf}
\end{align}

\subsubsection{Interference interpretation}\label{intinterp}

\begin{figure}[htb!]
\begin{overpic}[height=5.5cm,width=.22\linewidth]{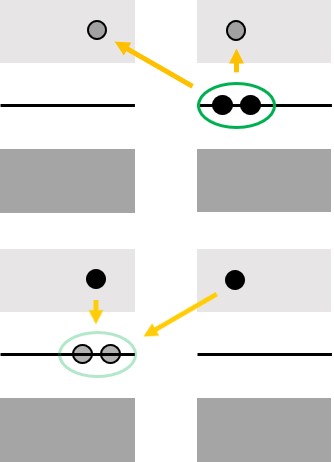}%
 \put (16,105) {\normalsize	(a)}
\end{overpic}\hspace*{20mm}
\begin{overpic}[height=5.5cm,width=.22\linewidth]{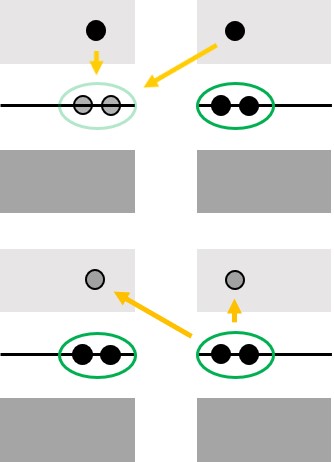}%
 \put (16,105) {\normalsize	(b)}
\end{overpic}\hspace*{20mm}
\begin{overpic}[height=5.5cm,width=.22\linewidth]{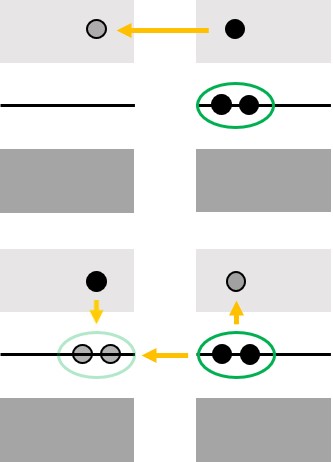}%
 \put (16,105) {\normalsize	(c)}
\end{overpic}
\caption{Schematic of the microscopic processes contributing to the pair current. The translucent elements denote that the process happens in the next step. For instance, in (a), the translucent pair in the lower panel will form eventually in the subsequent step once the two electron-like quasiparticles pair up. (a) The dominant process for low temperatures, wherein a Cooper pair breaks up into two Bogoliubov quasiparticles which combine on the other side of the junction. (b) Another version of (a), which is sub-dominant as it hinges upon the presence of two thermally excited quasiparticles. (c) Unlike (b), this one requires only one thermally excited quasiparticle.}
\label{FigS1}
\end{figure}
For a physical interpretation of the terms entering the expression for the current, we refer back to Eq.\eqref{Isstemp}. We look at the pair current, given by its second line, for a simple discussion. The first term, $e^{-i\frac{(\phi(t)+\phi(t_1))}{2}} \tilde{f}_{R,1,R,1,\sigma\sigma'}^{-+}(t,t_1)f_{L,1,L,1,\sigma'\sigma}^{+-}(t_1,t)$, represents the tunneling from the right to the left lead. Defining $|i\rangle=|g_L\rangle |g_R\rangle$ as the initial ground state, which is a product of the BCS ground states of each lead, a single tunneling event at time $t_1$ yields $|i'\rangle= e^{-i\frac{\phi(t_1)}{2}}c^\dagger_{L,\sigma'}(t_1)c_{R,\sigma'}(t_1)|g_L\rangle |g_R\rangle$. Its time evolved version at the present time $t$ is $|f\rangle= e^{+i\frac{\phi(t)}{2}}c_{L,\sigma}(t)c_{R,\sigma}^\dagger(t)|g_L+2\rangle |g_R-2\rangle$, where $\pm 2e$ denotes the addition/removal of a Cooper pair. Note that the sign of the phase changes as two electrons are transferred. The Cooper instability allows the exchange of a pair without any energy cost. Note that we have the opposite spin $\sigma'\neq \sigma$ because $c_{R,\sigma}^\dagger|g_R-2\rangle=c_{R,\sigma'}|g_R\rangle$ as the pair contains an electron with $\sigma$ and another with $\sigma'\neq \sigma$. Finally, the corresponding current is obtained as the overlap of these two states,
\begin{align}
I^S_{R\to L}\sim\langle f|i'\rangle=&e^{-i\frac{(\phi(t)+\phi(t_1))}{2}}\langle g_L+2|\langle g_R-2|  c_{R,\sigma}(t)c_{L,\sigma}^\dagger(t)c_{L,\sigma'}^\dagger(t_1)c_{R,\sigma'}(t_1)|g_L\rangle |g_R\rangle\nonumber\\
=&e^{-i\frac{(\phi(t)+\phi(t_1))}{2}}\tilde{f}_{R,1,R,1,\sigma\sigma'}^{-+}(t,t_1)f_{L,1,L,1,\sigma'\sigma}^{+-}(t_1,t),
\end{align}
which effectively represents the transfer of a Cooper pair proceeding via intermediate quasiparticle excitations (IQPs). These IQPs tunnel at different times $t,t_1$, which leads to interference due to their phases being different. These include not only the phases accumulated by the quasiparticles due to the bias, but also the dynamic phases representing the time-evolution of the quasiparticles, with the latter being accounted for by the bare Green's functions. The former depends only on the time spent in the biased lead and hence, for the quasiparticles having already tunneled at time $t_1$, there is no further change.

A similar analysis may be done for the normal current where, instead of exchanging a pair, the final and initial ground states are the same.

\subsubsection{Phenomenological picture of current oscillations}

\paragraph{Pair current:--}
We denote the BCS ground state of the lead $j$ as $|g_j\rangle$. Given the initial product ground state $|\psi\rangle(t_0)=|g_L,g_R\rangle$ of the combined left(L)-right(R) lead junction, pair tunneling generates the states $|g_L+2, g_R-2\rangle$ and $|g_L-2, g_R+2\rangle$ denoting a pair transferred to the left and the right side relative to the initial state, respectively. In the interaction picture, the single-particle tunneling Hamiltonian in Eq. \eqref{HT} becomes $\tilde{H}_T=\sum_\sigma -\mathcal{T}e^{iH_0 t}\big(e^{-i\phi(t)/2}c_{L,1\sigma}^\dagger c_{R,1\sigma}+e^{i\phi(t)/2}\hat{c}_{R,1\sigma}^\dagger c_{L,1\sigma}\big)e^{-iH_0 t}$. Thus, we obtain,
\begin{align}
|\psi\rangle(t)=&\psi_0(t)|g_L, g_R\rangle+\psi_{R\to L}(t)|g_L+2, g_R-2\rangle + \psi_{L\to R}(t)|g_L-2, g_R+2\rangle,
\end{align}
with 
\begin{subequations}
\begin{align}
\psi_{0}(t)=&1+\mathcal{O}(\mathcal{T}^2)\\
\psi_{R\to L}(t)=&\psi'+\frac{1}{(i\hbar)^2}\int_{t_0}^tdt'\int_{t_0}^{t'}dt'' \sum_{\epsilon_R,\epsilon_L}\langle g_L+2, g_R-2| \tilde{H}_T(t')|\epsilon_L\rangle|\epsilon_R\rangle \langle \epsilon_L|\langle \epsilon_R|\tilde{H}_T(t'')|g_L,g_R\rangle\hspace*{2mm}\nonumber\\
=&\psi'+\frac{1}{(i\hbar)^2}\mathcal{T}^2\int_{t_0}^tdt'\int_{t_0}^{t'}dt'' \sum_{\epsilon_R,\epsilon_L}e^{-i\frac{\phi(t')}{2}-i\frac{\phi(t'')}{2}}  e^{-i(\epsilon_L+\epsilon_R)t'}e^{+i(\epsilon_L+\epsilon_R)t''}\\
\psi_{L\to R}(t)=&\psi'+\frac{1}{(i\hbar)^2}\int_{t_0}^tdt'\int_{t_0}^{t'}dt'' \sum_{\epsilon_R,\epsilon_L}\langle g_L-2, g_R+2| \tilde{H}_T(t')|\epsilon_L\rangle|\epsilon_R\rangle \langle \epsilon_L|\langle \epsilon_R|\tilde{H}_T(t'')|g_L,g_R\rangle\hspace*{2mm} \nonumber\\
=&\psi'+\frac{1}{(i\hbar)^2}\mathcal{T}^2\int_{t_0}^tdt'\int_{t_0}^{t'}dt'' \sum_{\epsilon_R,\epsilon_L}e^{i\frac{\phi(t')}{2}+i\frac{\phi(t'')}{2}}  e^{-i(\epsilon_L+\epsilon_R)t'}e^{+i(\epsilon_L+\epsilon_R)t''}
\end{align}
\end{subequations}
The factor $\psi'$ reflects the fact that the initial uncoupled BCS ground states already contain a superposition of states with different numbers of pairs in each lead. Note that for the same set of states $\epsilon_{L,R}$, the factor $\psi'$ is same for both $\psi_{R\to L}$ and $\psi_{L\to R}$ as we consider the same superconductors on both sides of the junction. The transition $|g_L, g_R\rangle\to|g_L-2, g_R+2\rangle$ is depicted in the process shown in Fig. \ref{FigS1}(b). In a perturbative calculation, states with more transferred pairs are higher order in the tunnel coupling. In this case, even though we have tunneling of Cooper pairs, the fact that in a perturbative sense it proceeds through intermediate quasiparticles provides an avenue for them to interfere amongst themselves. As shown in Fig. \ref{FigS1}(a), and from the expression above, there are two tunneling events involving the virtually excited quasiparticles occurring at different times, $t',t$. Since these quasiparticles can only accumulate the bias-dependent phase as long as they are in the biased (left) lead, they tunnel at different times with different phases, resulting in a time-dependent interference signature. This directly manifests as oscillations in the Josephson current. Noting that the IQPs injected from a lead share the same distribution as that lead~\cite{sSinghal2020}, the
current is obtained as,
\begin{align}
I\sim&\iint_{0}^\infty d\epsilon_L d\epsilon_R A^S(\epsilon_L)A^S(\epsilon_R)  \Re\bigg( \frac{\partial}{\partial t} |\psi_{R\to L}(t)|^2 -\frac{\partial}{\partial t} |\psi_{L\to R}(t)|^2 \bigg)( (1-f(\epsilon_L))(1-f(\epsilon_R))-f(\epsilon_L)f(\epsilon_R) )\nonumber\\
\sim&\mathcal{T}^2\iint_{0}^\infty d\epsilon_L d\epsilon_R A^S(\epsilon_L)A^S(\epsilon_R)  \Re\int_{-\infty}^{t}dt''e^{-i\frac{\phi(t)+\phi(t'')}{2}}  e^{-i(\epsilon_L+\epsilon_R)(t-t'')}( (1-f(\epsilon_L))(1-f(\epsilon_R))-f(\epsilon_L)f(\epsilon_R) )+\mathcal{O}(T^4)\nonumber\\
=&\mathcal{T}^2\iint_{0}^\infty d\epsilon_L d\epsilon_R A^S(\epsilon_L)A^S(\epsilon_R)  \Re\int_{-\infty}^{t}dt''e^{-i\frac{\phi(t)+\phi(t'')}{2}}  e^{-i(\epsilon_L+\epsilon_R)(t-t'')}( 1-f(\epsilon_L)-f(\epsilon_R) ).
\end{align}
Note that now we have the anomalous spectral functions $A^S(\epsilon)=(\Delta/\epsilon)A^N(\epsilon)$ which represent the quasiparticle excitation distribution generated by pair breaking processes. Regarding the group of terms making up the factor with the Fermi functions, in the first term, the Fermi functions are chosen so as to reflect the requirement that both the states $\epsilon_{L,R}$ must be empty to facilitate the pair tunneling. In the second term, we consider an equivalent process where both the quasiparticles are initially occupied, as shown in Fig. \ref{FigS1}(b), with the minus sign arising due to the fermionic anti-commutation as the same transport happens in the opposite order. While the latter is suppressed at low temperatures, it nevertheless yields the correct expression which may be generalised to finite temperatures. There are also additional distinct processes, such as the one shown in Fig. \ref{FigS1}(c), which do not survive at zero-temperature. They may alternatively be obtained from Eq. \eqref{knksgen1} by splitting the range of the $\epsilon_{L,R}$ integrals into positive and negative sectors, which we omit here for brevity. Hence, we obtain,
\begin{align}
I\sim&\Re \mathcal{T}^2\int_{-\infty}^{t}dt'e^{-i\frac{\phi(t)+\phi(t')}{2}} \bigg(\underbrace{\iint_{-\infty}^\infty d\epsilon_L d\epsilon_R A^S(\epsilon_L)A^S(\epsilon_R)   e^{-i(\epsilon_L-\epsilon_R)(t-t')}( f(\epsilon_L)-f(\epsilon_R) )}_{=iK_S(t-t')}\bigg)\label{Isphenom1}\\
=&\Im \mathcal{T}^2e^{-i\phi(t)} \bigg( \underbrace{\int_{-\infty}^{t}dt'e^{i\frac{\phi(t)-\phi(t')}{2}}\Im\iint_{-\infty}^\infty d\epsilon_L d\epsilon_R A^S(\epsilon_L)A^S(\epsilon_R)   e^{-i(\epsilon_L-\epsilon_R)(t-t')}( f(\epsilon_L)-f(\epsilon_R) )}_{J_p(t)}\bigg).
\end{align}
Eq. \eqref{Isphenom1} resembles Eq. \eqref{IQt} along with Eq. \eqref{knksgen1}.

\paragraph{Normal current:--}

We apply the voltage $V(t)$ to the left lead. Consider two quasiparticles at energies $\epsilon_{L}$ and $\epsilon_{R}$ with the corresponding states $|L\rangle$ and $|R\rangle$ in the left and right leads, respectively. 
Their time evolution is given by,
\begin{align}
i\frac{\partial}{\partial t}\begin{bmatrix}
|R\rangle\\ |L\rangle
\end{bmatrix}=&\begin{bmatrix}
\epsilon_R & \mathcal{T} \\ \mathcal{T} & \epsilon_L+V(t)
\end{bmatrix} \begin{bmatrix}
|R\rangle \\ |L\rangle
\end{bmatrix}.
\end{align}
Starting first with the case where the state $|L\rangle$ is occupied while $\psi_R$ is unoccupied, with $|\psi\rangle(t_0)=|L\rangle$, we have to order $\mathcal{T}$,
\begin{align}
|\psi(t)\rangle=&\psi_L(t)|L\rangle+\psi_R(t)|R\rangle=\Big(e^{-i\epsilon_L (t-t_0)}\Big)|L\rangle+\Big(-ie^{-i\epsilon_R (t-t_0)}\int_{t_0}^t dt' e^{i\epsilon_R (t'-t_0)}\mathcal{T}e^{-i\epsilon_L (t'-t_0)}e^{-i\int_{t_0}^{t'}d\tau V(\tau)}\Big)|R\rangle
\end{align}
The term with $|R\rangle$ contains an integral over $t'$, representing the moment when the tunneling event occurred. As such, for a given time $t'$, the bias-dependent phase is fixed at $\phi(t')=\int_{t_0}^{t'}d\tau V(\tau)$, corresponding to the phase picked up by $|L\rangle$ in the biased left lead until the tunneling event. As we derive below, this leads to interference between states having tunneled at different times, which in turn leads to oscillations in the current as we mentioned in the main text.

The $L\to R$ current may be obtained from the time-derivative of the quasiparticle density on the $R$ side, summed over all possible combinations of occupied states $\epsilon_L$ and unoccupied states $\epsilon_R$,
\begin{align}
I_{L\to R}\sim&\iint_{-\infty}^\infty d\epsilon_L d\epsilon_R A^N(\epsilon_L)A^N(\epsilon_R)  \frac{\partial}{\partial t} |\psi_R(t)|^2f(\epsilon_L)(1-f(\epsilon_R))\nonumber\\
=&\iint_{-\infty}^\infty d\epsilon_L d\epsilon_R A^N(\epsilon_L)A^N(\epsilon_R) 2\Re\ \mathcal{T}^2 e^{-i(\epsilon_R-\epsilon_L)t} e^{i\phi(t)} \int_{-\infty}^t dt' e^{-i\phi(t')+i(\epsilon_R-\epsilon_L)t'} f(\epsilon_L)(1-f(\epsilon_R))\nonumber\\
=&\iint_{-\infty}^\infty d\epsilon_L d\epsilon_R A^N(\epsilon_L)A^N(\epsilon_R) 2\Re\ \mathcal{T}^2 e^{i\phi(t)} \int_{0}^\infty d\tau e^{-i\phi(t-\tau)-i(\epsilon_R-\epsilon_L)\tau} f(\epsilon_L)(1-f(\epsilon_R))
\end{align}
where we have taken $t_0\to-\infty$. Here $\nu(\epsilon)$ denotes the density of states. Note that the same result is obtained on using the time-derivative of $|\tilde{\psi}_L|^2$ instead. The Fermi functions have been added to reflect the fact that for this process to happen, $\epsilon_L$ must be initially occupied while $\epsilon_R$ must be unoccupied. 

Similarly, $I_{R\to L}$ is obtained by considering initially an occupied state $\epsilon_R$ in right lead with $\psi_R(t_0)=1$ and an initially unoccupied state $\epsilon_L$ in the left lead with $\psi_L(t_0)=0$.
\begin{align}
I_{R\to L}\sim&\iint_{-\infty}^\infty d\epsilon_L d\epsilon_R A^N(\epsilon_L)A^N(\epsilon_R)  \frac{\partial}{\partial t} |\psi_L(t)|^2f_R(\epsilon_R)(1-f_L(\epsilon_L))\nonumber\\
\approx&\iint_{-\infty}^\infty d\epsilon_L d\epsilon_R A^N(\epsilon_L)A^N(\epsilon_R) 2\Re\ \mathcal{T}^2 e^{-i\phi(t)} \int_{0}^\infty d\tau e^{i\phi(t-\tau)+i(\epsilon_R-\epsilon_L)\tau} f(\epsilon_R)(1-f(\epsilon_L)).
\end{align}
The net current is thus obtained as,
\begin{align}
I=L_{L\to R}-I_{R\to L}\sim& 2 \mathcal{T}^2 \Re\ e^{i\phi(t)} \int_{-\infty}^\infty dt' e^{-i\phi(t')}\bigg(\Theta(t-t')\underbrace{\iint_{-\infty}^\infty d\epsilon_L d\epsilon_R A^N(\epsilon_L)A^N(\epsilon_R)\ e^{-i(\epsilon_R-\epsilon_L)(t-t')} (f(\epsilon_R)-f(\epsilon_L))}_{iK_N(t-t')}\bigg)\label{phenoI}
\end{align}
Following the steps outlined in the previous section, the Werthamer current reduces to the same form, as evident from Eq. \eqref{knksgen1}. This analysis clearly reveals that the acceleration of the intermediate quasiparticles constituting the current by the bias is the primary source of the current oscillations. \\

\subsubsection{Kernels}
\begin{figure}[htb!]
\includegraphics[height=5.5cm,width=.65\linewidth]{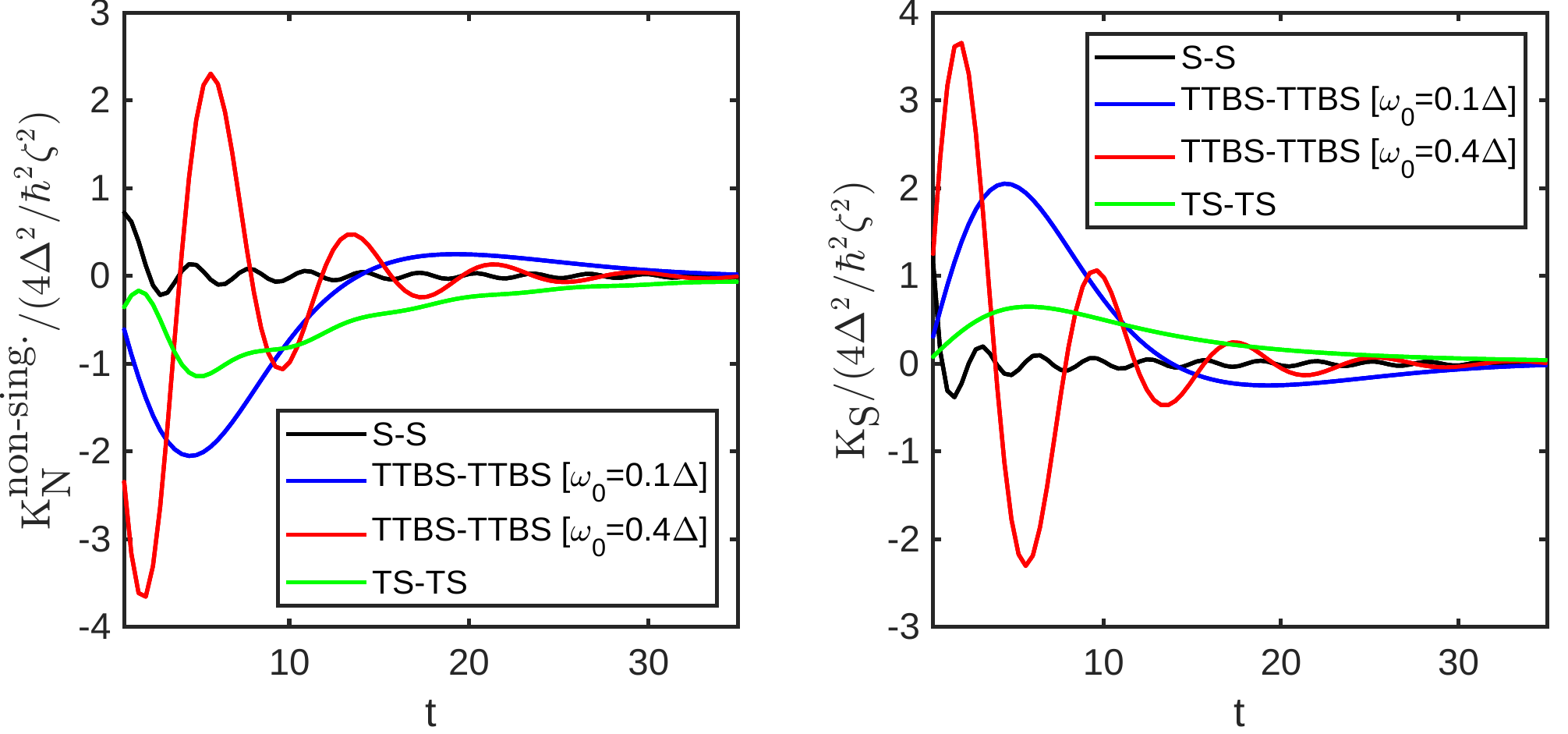}
\caption{The kernels for the various cases considered in this work. Band contributions are neglected in the case with bound states (BS) as they do not lead to NFJE. For $K_N$ we only show the non-singular non-Ohmic part of the kernel to highlight the retardation. We have used $\Gamma=0.1\Delta$. The cases with bound states not only have a longer decay/memory time $\sim\hbar/\Gamma$, but they also show oscillations commensurate with the bound state frequency $\omega_0$.}
\label{FigS2}
\end{figure}
We obtain the kernels $K_N,K_S$ following Refs.~\cite{sHarris1975a,sAverin2021}. The pair kernel is given by, 
\begin{align}
K_S(t)=& \Im \bigg\{\iint_{-\infty}^\infty\frac{dr}{2\pi}\frac{ds}{2\pi}e^{ir t}e^{-is t}B_L(s)B_R(r)\big[f(r)-f(s)\big]\bigg\},\quad \big(r=\Omega-\omega/2,\hspace*{2mm}s=\Omega+\omega/2\big)\nonumber\\
=& \frac{\Delta_L\Delta_R}{2\pi^2\hbar^2}\bigg(\Big[\int_0^\infty da \cos\Big(\frac{\Delta_L ta}{\hbar}\Big)B_L\Big(\frac{\Delta_L a}{\hbar}\Big)\Big]\Big[\int_0^\infty db \sin\Big(\frac{\Delta_R tb}{\hbar}\Big)B_R\Big(\frac{\Delta_R b}{\hbar}\Big)\Big]\nonumber\\
&\hspace*{9.5mm}+\Big[\int_0^\infty da \sin\Big(\frac{\Delta_L ta}{\hbar}\Big)B_L\Big(\frac{\Delta_L a}{\hbar}\Big) \Big]\Big[\int_0^\infty db \cos\Big(\frac{\Delta_R tb}{\hbar}\Big)B_R\Big(\frac{\Delta_R b}{\hbar}\Big)\Big]\bigg),\quad\big(r=\Delta_R a/\hbar,\hspace*{2mm}s=\Delta_L b/\hbar\big),\nonumber\\
=& -\frac{2\Delta_L\Delta_R}{\hbar^2\zeta^2}\Big[ J_0\Big(\frac{\Delta_L}{\hbar}t\Big)Y_0\Big(\frac{\Delta_R}{\hbar}t\Big)+Y_0\Big(\frac{\Delta_L}{\hbar}t\Big)J_0\Big(\frac{\Delta_R}{\hbar}t\Big) \Big].\label{Ksss}
\end{align}
The quasiparticle kernel is obtained similarly by separating the Ohmic part,
\begin{align}
K_N(t)=&-\frac{2\Delta_L\Delta_R}{\hbar^2\zeta^2}\Big[ J_1\Big(\frac{\Delta_L}{\hbar}t\Big)Y_1\Big(\frac{\Delta_R}{\hbar}t\Big)+Y_1\Big(\frac{\Delta_L}{\hbar}t\Big)J_1\Big(\frac{\Delta_R}{\hbar}t\Big) \Big]+\frac{8}{\pi\zeta^2}\frac{d\delta(t)}{dt},\label{Knss}
\end{align}
These kernels oscillate at the gap frequency $2\Delta/\hbar$.

\paragraph{Bound states: }We use the exponential regularisation for the bound state spectral peak,
\begin{align}
\frac{4\Delta}{\zeta}\pi \delta(\hbar(\omega-\omega_0))&\to \frac{4\Delta}{\zeta}\pi  \bigg[\frac{1}{2\Gamma}\exp\bigg(\frac{-\hbar|\omega-\omega_0|}{\Gamma}\bigg)\bigg].
\end{align}
Considering bound states located at $\omega_{0,L}$ and $\omega_{0,R}$ with spectral weights $h_{N,S}^L$ and $h_{N,S}^R$ in the left and the right leads, respectively, and following the same procedure as before, the kernels are obtained as, 
\begin{align}
K_S(t)=&\frac{2\Delta_L\Delta_R}{\hbar^2\zeta^2}\bigg[ \bigg\{ h_S^L\frac{2\cos(\omega_{0,L}t)-e^{-\frac{\hbar\omega_{0,L}}{\Gamma}}}{1+(\frac{\Gamma t}{\hbar})^2}\bigg\}\bigg\{ h_S^R\frac{2\sin(\omega_{0,R}t)+\frac{\Gamma t}{\hbar}e^{-\frac{\hbar\omega_{0,R}}{\Gamma}}}{1+(\frac{\Gamma t}{\hbar})^2}\bigg\}+(L\leftrightarrow R)\bigg],\label{Ksssabsn}
\end{align}
\begin{align}
K_N(t)=& -\frac{2\Delta_L\Delta_R}{\hbar^2\zeta^2}\bigg[ \bigg\{ h_N^L\frac{2\cos(\omega_{0,L}t)-e^{-\frac{\hbar\omega_{0,L}}{\Gamma}}}{1+(\frac{\Gamma t}{\hbar})^2}\bigg\}\bigg\{ h_N^R\frac{2\sin(\omega_{0,R}t)+\frac{\Gamma t}{\hbar}e^{-\frac{\hbar\omega_{0,R}}{\Gamma}}}{1+(\frac{\Gamma t}{\hbar})^2}\bigg\}+(L\leftrightarrow R)\bigg]+\frac{8}{\pi\zeta^2}\frac{d\delta(t)}{dt}\label{Knaabsn}.
\end{align}
Note that, in the limit $\omega_{0,L/R}\to 0$, the bound state contribution becomes mathematically the same as its MZM counterpart, which we show later in Eqs. \eqref{Kststsn} and \eqref{Kntstsn}. Hence, the resulting NFJE bears the same functional form, as captured by Eq. \eqref{nfjemzm} in the main text.

\subsection{p-wave (Kitaev chain [TS]) superconductor}
\subsubsection{Degenerate states}
Prior to obtaining the expression for the current, we must establish a formalism to handle the ground state degeneracy in each TS lead. In this situation, we must start with suitable initial ground states to obtain a convergent and well-defined result from a perturbative calculation of the current~\cite{sBrouder2009}. Specifically, we must use one of the eigenstates $v_j$ $(H_T(t)v_j=E_j(t)v_j)$ of the projected tunneling Hamiltonian, $PH_TP$, where $P$ is the projector to the degenerate subspace of initial states. In our case, considering two isolated Kitaev chains, we write the tunneling Hamiltonian as,
\begin{align}
H_T(t)=&\mathcal{T}e^{-i\phi(t)/2}c_{LN}^\dagger c_{R1}+\mathcal{T}e^{i\phi(t)/2} c_{R1}^\dagger  c_{LN}.
\end{align}
In order to define the projected tunneling Hamiltonian $PH_TP$, we use the four-fold degenerate space of ground states of the two disconnected Kitaev chains: $|g_{Le}g_{Re}\rangle$, $|g_{Le}g_{Ro}\rangle$, $|g_{Lo}g_{Re}\rangle$, $|g_{Lo}g_{Ro}\rangle$. Here $|g_{Lj}g_{Rk}\rangle$ denotes the product state of the left (L) chain being in $|g_j\rangle$ and the right (R) chain being in $|g_k\rangle$, with $j,k\in\{e,o\}$ denoting the even and odd parity ground states. Following the procedure mentioned above along with Eq. \eqref{Skcrot}, we obtain,
\begin{subequations}
\begin{align}
|v_1\rangle=&\frac{|g_{Le}g_{Re}\rangle+|g_{Lo}g_{Ro}\rangle}{\sqrt{2}}:\quad E_{v_1}=\mathcal{T}\cos\left(\frac{\phi(t)}{2}\right),\quad \mathpzc{p}_I=1\\
|v_2\rangle=&\frac{|g_{Le}g_{Re}\rangle-|g_{Lo}g_{Ro}\rangle}{\sqrt{2}}:\quad E_{v_1}=-\mathcal{T}\cos\left(\frac{\phi(t)}{2}\right),\quad \mathpzc{p}_I=-1\\
|v_3\rangle=&\frac{|g_{Le}g_{Ro}\rangle+|g_{Lo}g_{Re}\rangle}{\sqrt{2}}:\quad E_{v_1}=-\mathcal{T}\cos\left(\frac{\phi(t)}{2}\right),\quad \mathpzc{p}_I=1\\
|v_4\rangle=&\frac{|g_{Le}g_{Ro}\rangle-|g_{Lo}g_{Re}\rangle}{\sqrt{2}}:\quad E_{v_1}=\mathcal{T}\cos\left(\frac{\phi(t)}{2}\right),\quad \mathpzc{p}_I=-1,
\end{align}\label{bellgs}
\end{subequations}
with $\mathpzc{p}_I$ being the parity defined by the MZMs located near the junction. Note that, these are the four Bell states considering the two qubits formed by the two ground states of each lead. We assume that the junction stays in the ground state manifold, given by the adiabatically-evolved Gellmann-Low state formed from the linear combination of $|v_2\rangle$ and $|v_3\rangle$, prior to the application of the bias. It is also permissible to start entirely within the excited state manifold $|v_1\rangle$ and $|v_4\rangle$, however, we cannot start with a mixture of ground state and excited states within the perturbative scheme to obtain a convergent result~\cite{sBrouder2009}.

\subsubsection{Tunneling current calculation}
The steps are largely the same as those for the previously considered BCS Josephson junction. However, there is one fundamental difference: The mixing of the ground states between the two isolated leads establishes ``initial correlations" between them, as evident from the form of the states $|v_{1\ldots4}\rangle$. With such a non-trivial ground state having initial correlations, the average of operators, in general, do not factorise~\cite{sHall1975,sLeeuwen2012}. For instance, considering two kinds of fermion operators $c_{1,2}$, which in the present case could represent fermions from the two different leads, $\langle c^\dagger_1 c^\dagger_2c_2c_1\rangle=\langle c^\dagger_1 c_1\rangle\langle c^\dagger_2 c_2\rangle+\Lambda_4$, where $\Lambda_4$ is a two-particle vertex which destroys the factorisation. In our case, the non-vanishing contributions to $G_{R,1,L,1,j}^{+-}$, where $j$ denotes the chosen initial state $|v_j\rangle$, is obtained as~\cite{sCuevasbook2017,sJauhobook2008},
\begin{align}
&G_{R,1,L,1,j}^{+-}\left(t, t'\right)\nonumber\\
&=g_{R,1,L,1,j}^{+-}\left(t, t^{\prime}\right) \nonumber\\
&\hspace*{3.5mm}- \int_{-\infty}^\infty dt_1 (-\mathcal{T})\Big[ e^{-i\phi(t_1^+)/2}{\underbrace{\big\langle v_j\big|\mathbf{T}c_{R,1}(t^+)c_{L,1}^\dagger(t'^-)c_{L,1}^\dagger(t_1^+) c_{R,1}(t_1^+)\big|v_j\big\rangle}_{P1}}+e^{i\phi(t_1^+)/2}\big\langle v_j\big|\mathbf{T}c_{R,1}(t^+)c_{L,1}^\dagger(t'^-)c_{R,1}^\dagger(t_1^+) c_{L,1}(t_1^+) \Big] \big|v_j\big\rangle\Big]\nonumber\\
&\hspace*{3.5mm} + \int_{-\infty}^\infty dt_1(-\mathcal{T})\Big[e^{-i\phi(t_1^-)/2}{\underbrace{\big\langle v_j\big|\mathbf{T}c_{R,1}(t^+)c_{L,1}^\dagger(t'^-)c_{L,1}^\dagger(t_1^-) c_{R,1}(t_1^-)\big|v_j\big\rangle}_{P2}}+e^{i\phi(t_1^-)/2}\big\langle v_j\big|\mathbf{T}c_{R,1}(t^+)c_{L,1}^\dagger(t'^-)c_{R,1}^\dagger(t_1^-) c_{L,1}(t_1^-) \Big] \big|v_j\big\rangle\Big].\label{Grlcontraction}
\end{align}
The first term, $g_{R,1,L,1,j}^{+-}\left(t, t^{\prime}\right)$, arises due to the initial correlations. As we show next, it contributes to the parity-dependent standard Majorana-induced fractional Josephson current. Note that, in topologically trivial systems with a non-degenerate ground state, this term typically vanishes. While such anomalous parity-dependent terms also exist at order $\mathcal{T}$, fortunately they cancel out. This can be seen by carrying out the calculation in Eq. \eqref{Grlcontraction} using Eq. \eqref{Skcrot}. We demonstrate it only for the terms contributing to the pair-current (denoted $P1$ and $P2$ in Eq. \eqref{Grlcontraction}), for brevity. 
\begin{itemize}
\item The first term contributing to the pair current, $P1$, becomes,
\begin{align}
P1=&\big\langle \mathbf{T}c_{R,1}(t^+)c_{L,1}^\dagger(t'^-)c_{L,1}^\dagger(t_1^+) c_{R,1}(t_1^+) \big\rangle_{v_j}\nonumber\\
&= \sum_{j,k}v_1^k u_1^j v_1^j u_1^k \big\langle \mathbf{T}d_{R,k}(t^+)d_{L,j}(t'^-) d_{L,j}^\dagger(t_1^+) d^\dagger_{R,k}(t_1^+) \big\rangle_{v_j} + v_1^k v_1^j u_1^j u_1^k {\color{blue}\big\langle \mathbf{T}d_{R,k}(t^+)d_{L,j}^\dagger(t'^-) d_{L,j}(t_1^+) d^\dagger_{R,k}(t_1^+) \big\rangle_{v_j}} \nonumber\\
&\hspace*{2.5mm}+u_1^k u_1^j v_1^j v_1^k {\color{red}\big\langle \mathbf{T}d_{R,k}^\dagger(t^+)d_{L,j}(t'^-) d_{L,j}^\dagger(t_1^+) d_{R,k}(t_1^+) \big\rangle_{v_j}} + u_1^k v_1^j u_1^j v_1^k {\color{green}\big\langle \mathbf{T}d_{R,k}^\dagger(t^+)d_{L,j}^\dagger(t'^-) d_{L,j}(t_1^+) d_{R,k}(t_1^+) \big\rangle_{v_j}}\label{P1}
\end{align}
In this section we denote $v_j^1=\frac{\psi_{L,j}+\psi_{R,j}}{\sqrt{2}}$ and $u_j^1=\frac{\psi_{L,j}-\psi_{R,j}}{\sqrt{2}}$ for brevity. Looking at the first term in Eq. \eqref{P1} (named $P1-$black as it's written in black) here, we can separate each $d_j,d_j^\dagger$ operator into the MZM $(j=1)$ and non-MZM contribution $(j\neq 1)$ to obtain,
\begin{align}
P_1-\text{black}=&\sum_{j,k}\big\langle \mathbf{T}d_{R,k}(t^+)d_{L,j}(t'^-) d_{L,j}^\dagger(t_1^+) d^\dagger_{R,k}(t_1^+) \big\rangle_{v_j}\nonumber\\
=&\sum_{j,k\neq 1}\big\langle \mathbf{T}d_{R,k}(t^+)d^\dagger_{R,k}(t_1^+)\big\rangle_{v_j}\big\langle\mathbf{T}d_{L,j}(t'^-) d_{L,j}^\dagger(t_1^+)  \big\rangle_{v_j}\nonumber\\
&+\underbrace{\Big[ \big\langle \mathbf{T}d_{R,k=1}(t^+)d_{L,j=1}(t'^-) d_{L,j= 1}^\dagger(t_1^+) d^\dagger_{R,k=1}(t_1^+) \big\rangle_{v_j}- \big\langle \mathbf{T}d_{R,k=1}(t^+) d^\dagger_{R,k=1}(t_1^+) \big\rangle_{v_j}\big\langle\mathbf{T}d_{L,j=1}(t'^-) d_{L,j= 1}^\dagger(t_1^+)\big\rangle_{v_j}\Big]}_{\Lambda_4^{P1-\text{black}}}.
\end{align}
Since we are restricting ourselves to order $\mathcal{T}^2$ while calculating the current, and consequently to order $\mathcal{T}$ while calculating $G_{R,1,L,1,j}^{+-}$, we only require $\Lambda_4$ at $\mathcal{O}(T^0)$. This is evaluated as,
\begin{align}
\Lambda_4^{P1-\text{black}}=&\big\langle \mathbf{T}d_{R,k=1}(t^+)d_{L,j=1}(t'^-) d_{L,j= 1}^\dagger(t_1^+) d^\dagger_{R,k=1}(t_1^+) \big\rangle_{v_j}- \big\langle \mathbf{T}d_{R,k=1}(t^+) d^\dagger_{R,k=1}(t_1^+) \big\rangle_{v_j}\big\langle\mathbf{T}d_{L,j=1}(t'^-) d_{L,j= 1}^\dagger(t_1^+)\big\rangle_{v_j} \nonumber\\
=&\begin{cases}
-\big\langle d_{L,j=1}(t'^-) d_{R,k=1}(t^+)d_{L,j= 1}^\dagger(t_1^+) d^\dagger_{R,k=1}(t_1^+) \big\rangle_{v_j}- \big\langle d_{R,k=1}(t^+) d^\dagger_{R,k=1}(t_1^+) \big\rangle_{v_j}\big\langle d_{L,j=1}(t'^-) d_{L,j= 1}^\dagger(t_1^+)\big\rangle_{v_j}\Big];\quad t>t_1\\
-\big\langle d_{L,j=1}(t'^-) d_{L,j= 1}^\dagger(t_1^+) d^\dagger_{R,k=1}(t_1^+)d_{R,k=1}(t^+) \big\rangle_{v_j}+ \big\langle d^\dagger_{R,k=1}(t_1^+)d_{R,k=1}(t^+)  \big\rangle_{v_j}\big\langle d_{L,j=1}(t'^-) d_{L,j= 1}^\dagger(t_1^+)\big\rangle_{v_j}\Big]; \quad t<t_1
\end{cases}\nonumber\\
=&\begin{cases}
\frac{1}{4};\quad |v_j\rangle=|v_1\rangle,|v_2\rangle\\
-\frac{1}{4};\quad |v_j\rangle=|v_3\rangle,|v_4\rangle
\end{cases}
\end{align}
\item  Now, the second term contributing to the pair current, $P2$ in Eq. \eqref{Grlcontraction}, is evaluated as,
\begin{align}
P2=&\big\langle \mathbf{T}c_{R,1}(t^+)c_{L,1}^\dagger(t'^-)c_{L,1}^\dagger(t_1^-) c_{R,1}(t_1^-) \big\rangle_{v_j}\nonumber\\
&=\sum_{j,k} v_1^k u_1^j v_1^j u_1^k \big\langle \mathbf{T}d_{R,k}(t^+)d_{L,j}(t'^-) d_{L,j}^\dagger(t_1^-) d^\dagger_{R,k}(t_1^-) \big\rangle_{v_j} + v_1^k v_1^j u_1^j u_1^k {\color{blue}\big\langle \mathbf{T}d_{R,k}(t^+)d_{L,j}^\dagger(t'^-) d_{L,j}(t_1^-) d^\dagger_{R,k}(t_1^-) \big\rangle_{v_j}} \nonumber\\
&\hspace*{2.5mm}+u_1^k u_1^j v_1^j v_1^k {\color{red}\big\langle \mathbf{T}d_{R,k}^\dagger(t^+)d_{L,j}(t'^-) d_{L,j}^\dagger(t_1^-) d_{R,k}(t_1^-) \big\rangle_{v_j}} + u_1^k v_1^j u_1^j v_1^k {\color{green}\big\langle \mathbf{T}d_{R,k}^\dagger(t^+)d_{L,j}^\dagger(t'^-) d_{L,j}(t_1^-) d_{R,k}(t_1^-) \big\rangle_{v_j}}.\label{P2}
\end{align}
Looking at the first term in Eq. \eqref{P2} ($P2-$black) and on separating the operators as before,
\begin{align}
P_2-\text{black}=&\sum_{j,k}\big\langle \mathbf{T}d_{R,k}(t^+)d_{L,j}(t'^-) d_{L,j}^\dagger(t_1^-) d^\dagger_{R,k}(t_1^-) \big\rangle_{v_j}\nonumber\\
=&\sum_{j,k\neq 1}\big\langle \mathbf{T}d_{R,k}(t^+)d^\dagger_{R,k}(t_1^-)\big\rangle_{v_j}\big\langle\mathbf{T}d_{L,j}(t'^-) d_{L,j}^\dagger(t_1^-)  \big\rangle_{v_j}\nonumber\\
&+\underbrace{\Big[ \big\langle \mathbf{T}d_{R,k=1}(t^+)d_{L,j=1}(t'^-) d_{L,j= 1}^\dagger(t_1^-) d^\dagger_{R,k=1}(t_1^-) \big\rangle_{v_j}- \big\langle \mathbf{T}d_{R,k=1}(t^+) d^\dagger_{R,k=1}(t_1^-) \big\rangle_{v_j}\big\langle\mathbf{T}d_{L,j=1}(t'^-) d_{L,j= 1}^\dagger(t_1^-)\big\rangle_{v_j}\Big]}_{\Lambda_4^{P2-\text{black}}}
\end{align}
The vertex is turns out to be the same as the one for $P1-$black,
\begin{align}
\Lambda_4^{P2-\text{black}}=&\big\langle \mathbf{T}d_{R,k=1}(t^+)d_{L,j=1}(t'^-) d_{L,j= 1}^\dagger(t_1^-) d^\dagger_{R,k=1}(t_1^-) \big\rangle_{v_j}- \big\langle \mathbf{T}d_{R,k=1}(t^+) d^\dagger_{R,k=1}(t_1^-) \big\rangle_{v_j}\big\langle\mathbf{T}d_{L,j=1}(t'^-) d_{L,j= 1}^\dagger(t_1^-)\big\rangle_{v_j}\nonumber\\
=&\begin{cases}
\frac{1}{4};\quad |v_j\rangle=|v_1\rangle,|v_2\rangle\\
-\frac{1}{4};\quad |v_j\rangle=|v_3\rangle,|v_4\rangle
\end{cases}=\Lambda_4^{P1-\text{black}}.
\end{align} 
\end{itemize}
Since the two terms $P1$ and $P2$ contributing to the pair-current (marked in Eq. \eqref{Grlcontraction}) appear with opposite signs, corresponding to them lying on the forward and backward propagating branches of the Keldysh contour, the vertices cancel between $P1-$black and $P2-$black. Similarly, the vertices also cancel between $P1-$blue and $P2-$blue, and so on, yielding a factorised result,
\begin{align}
P1-P2=&\sum_{j,k} v_1^k u_1^j v_1^j u_1^k \Big(\big\langle \mathbf{T}d_{R,k}(t^+)d_{L,j}(t'^-) d_{L,j}^\dagger(t_1^+) d^\dagger_{R,k}(t_1^+) \big\rangle_{v_j}-\big\langle \mathbf{T}d_{R,k}(t^+)d_{L,j}(t'^-) d_{L,j}^\dagger(t_1^-) d^\dagger_{R,k}(t_1^-) \big\rangle_{v_j}\Big)+\ldots\nonumber\\
=&\Big( \sum_{j,k\neq 1} v_1^k u_1^j v_1^j u_1^k \big\langle \mathbf{T}d_{R,k}(t^+)d^\dagger_{R,k}(t_1^+)\big\rangle_{v_j}\big\langle\mathbf{T}d_{L,j}(t'^-) d_{L,j}^\dagger(t_1^+)  \big\rangle_{v_j}+\cancel{v_1^1 u_1^1 v_1^1 u_1^1\Lambda_4^{P1-\text{black}}}\nonumber\\
&-\sum_{j,k\neq 1} v_1^k u_1^j v_1^j u_1^k \big\langle \mathbf{T}d_{R,k}(t^+)d^\dagger_{R,k}(t_1^-)\big\rangle_{v_j}\big\langle\mathbf{T}d_{L,j}(t'^-) d_{L,j}^\dagger(t_1^-)  \big\rangle_{v_j}-\underbrace{\cancel{v_1^1 u_1^1 v_1^1 u_1^1\Lambda_4^{P2-\text{black}}}}_{=v_1^1 u_1^1 v_1^1 u_1^1\Lambda_4^{P1-\text{black}}}\Big)+\ldots\\
=&\big\langle \mathbf{T}c_{R,1}(t^+) c_{R,1}(t_1^+) \big\rangle_{v_j}\big\langle c_{L,1}^\dagger(t'^-)c_{L,1}^\dagger(t_1^+) \big\rangle_{v_j}-\big\langle \mathbf{T}c_{R,1}(t^+) c_{R,1}(t_1^-)\big\rangle_{v_j}\big\langle c_{L,1}^\dagger(t'^-)c_{L,1}^\dagger(t_1^-)\big\rangle_{v_j}.
\end{align}
Similarly, all the vertices cancel also for the normal current. Therefore, the factorisation holds at order $\mathcal{T}$. Note that while we have shown this factorisation for a single state $|v_j\rangle$, following the same steps as above it is easy to see that we can just as well start with a linear combination of the states in the ground state manifold $\{|v_2\rangle, |v_3\rangle\}$. 

Hence,
\begin{align}
G_{R,1,L,1,j}^{+-}\left(t, t'\right)=&g_{R,1,L,1,j}^{+-}\left(t, t^{\prime}\right) + \int_{-\infty}^\infty dt_1 (-\mathcal{T})\Theta(t-t_1)e^{i\phi(t_1)/2}\Big[ g_{R,1,j}^{-+}(t,t_1)g_{L,1,j}^{+-}(t_1,t')-g_{R,1,j}^{+-}(t,t_1)g_{L,1,j}^{-+}(t_1,t')\Big]\nonumber\\
&\hspace*{21.7mm} - \int_{-\infty}^\infty dt_1 (-\mathcal{T})\Theta(t-t_1)e^{-i\phi(t_1)/2}\Big[\tilde{f}_{R,1,j}^{-+}(t,t_1)f_{L,1,j}^{+-}(t_1,t')-\tilde{f}_{R,1,j}^{+-}(t,t_1)f_{L,1,j}^{-+}(t_1,t')\Big].\label{Grlfinal}
\end{align}

Following the steps employed previously for the s-wave junction, we obtain,
\begin{align}
I_j=&\frac{e(-\mathcal{T})}{\hbar}\Re \Big[e^{-i\frac{\phi(t)}{2}} g_{R,1,L,1,j}^{+-}\left(t, t\right)\Big]\nonumber\\
&+\frac{e(-\mathcal{T})^2}{\hbar}\int_{-\infty}^\infty dt_1 \Theta(t-t_1)\Re \Big[ e^{-i\frac{(\phi(t)-\phi(t_1))}{2}}\big(g_{R,1,j}^{-+}(t,t_1)g_{L,1,j}^{+-}(t_1,t)-g_{R,1,j}^{+-}(t,t_1)g_{L,1,j}^{-+}(t_1,t)\big)\nonumber\\
&\hspace*{44.5mm}-e^{-i\frac{(\phi(t)+\phi(t_1))}{2}}\big(\tilde{f}_{R,1,j}^{-+}(t,t_1)f_{L,1,j}^{+-}(t_1,t)-\tilde{f}_{R,1,j}^{+-}(t,t_1)f_{L,1,j}^{-+}(t_1,t)\big)\Big]\label{Itststemp}
\end{align}
The second term, proportional to $\mathcal{T}^2$, bears the same functional structure as in the s-wave BCS Josephson junction (Eq. \eqref{Isstemp}), although the constituent Green's functions correspond to a p-wave superconductor. This term results in NFJE. Note that, the corresponding physical mechanism is illustrated in Fig.\ref{Fig1} in the main text.

The current has an additional contribution $I^{(0)}\sim\mathcal{O}(\mathcal{T})$, given by the first term, which yields the standard Majorana-induced anomalous Josephson current~\cite{sKitaev2001,sLutchyn2010}. Using Eq. \eqref{Skcrot}, we have,
\begin{align}
I^{(0)}_j=&\frac{e(-\mathcal{T})}{\hbar}\Re \Big[e^{-i\frac{\phi(t)}{2}} g_{R,1,L,1,j}^{+-}\left(t, t^{\prime}\right)\Big]=\frac{e(\mathcal{T})}{\hbar}\sin\bigg(\frac{\phi(t)}{2}\bigg)\frac{\psi_{L,1}\psi_{R,N}}{2}\mathpzc{p}_j,\label{I0tststemp}
\end{align}
which gives a ground state dependent contribution, depending only on the parity of the inner MZMs, $\mathpzc{p}_j=(-1)^{n_{in}}$, formed by the two MZMs located at the ends of the wires near the junction. Note that physically, it doesn't proceed via the Bogoliubov excitations depicted in Fig.\ref{Fig1}. Instead it arises as the degenerate ground states are associated with different fermion parities and thus changes in ground state, which cost zero energy, can nevertheless contribute to the current.

\subsubsection{Parity dependence}
In this section we look at the ground state parity dependence of the current derived previously. Referring back to the interpretation of the terms entering the current as presented in Sec. \ref{intinterp}, we note that the $\mathcal{O}(\mathcal{T}^2)$ current involves the interference only between the quasiparticles which have already tunneled. This is evident from the fact that both the states $|i'\rangle$ and $|f\rangle$ introduced in Sec. \ref{intinterp} are the states \emph{after} the tunneling event. As such, even though the states $|i'\rangle$ and $|f\rangle$ are themselves dependent on the choice of the ground state and hence the parity, the interference amplitude which determines the current is independent of the parity as both $|i'\rangle$ and $|f\rangle$ depend on the parity in the exact same way. The crucial point is that the current is independent of the choice of the ground state. As such, starting in the ground state manifold $\{|v_2\rangle, |v_3\rangle\}$ as mentioned earlier, any parity-flipping process which alters the state $|v_i\rangle\to |v_{j\neq i}\rangle$ does not change the current. For the same reason, if such processes and enviromnental decoherence establish a statistically mixed ensemble of the states, such as that described by the density matrix $\rho=\sum_j w_j|v_j\rangle\langle v_j|$, the statistical Green's function~\cite{sBrouder2009} can be used to obtain the current. Since the NFJE current is independent of the ground state, they do not cancel. On the other hand, $I^{(0)}$ depends on parity and hence, the contributions from each state can cancel each other out.

On the contrary, in the case of $I^{(0)}$, a similar argument as in Sec. \ref{intinterp} may be repeated, but this time, the overlap is considered between the quasiparticle state having tunneled presently and the \emph{initial untouched ground state} without any tunneling events. As such, it explicitly probes the capacity of the initial ground state to let an electron tunnel. The structure of the ground states $|v_j\rangle$, which establishes entanglement between the even and odd parity states of the two leads, explicitly permits quasiparticle tunneling while still remaining within the ground state. Since the form of the ground states differ depending on the choice of the ground state, the resulting overlap and thus the current is parity dependent.

\subsubsection{Kernels}
We use the exponential regularisation for the MZM spectral peak,
\begin{align}
\frac{4\Delta}{\zeta}\pi \delta(\hbar\omega)&\to \frac{4\Delta}{\zeta}\pi  \bigg[\frac{1}{2\Gamma}\exp\bigg(\frac{-|\hbar\omega|}{\Gamma}\bigg)\bigg].
\end{align}
As before, the kernels are obtained as, 
\begin{align}
K_S(t)=&\frac{4\Delta_L\Delta_R}{\zeta^2\hbar^2}\frac{1}{1+(\frac{\Gamma t}{\hbar})^2}\frac{\frac{\Gamma t}{\hbar}}{1+(\frac{\Gamma t}{\hbar})^2}.\label{Kststsn}
\end{align}
\begin{align}
K_N(t)=& -\frac{\Delta_L\Delta_R}{2\pi^2\hbar^2}\Big(\frac{4}{\zeta}\Big)^2\bigg[\bigg\{{\color{blue}\frac{\pi}{2}\frac{1}{1+(\frac{\Gamma t}{\hbar})^2}}-\frac{\pi}{2}\bigg[1+J_1\Big(\frac{\Delta_L}{\hbar}t\Big)\bigg]+\frac{\pi}{4}\Big(\frac{\Delta_L}{\hbar}t\Big)\bigg[J_1\Big(\frac{\Delta_L}{\hbar}t\Big)\pi H_0\Big(\frac{\Delta_L}{\hbar}t\Big)+J_0\Big(\frac{\Delta_L}{\hbar}t\Big)\bigg(2-\pi H_1\Big(\frac{\Delta_L}{\hbar}t\Big)\bigg)\bigg]\bigg\}\nonumber\\
&\hspace*{14.75mm}\times\bigg\{\frac{\pi}{2}\frac{\hbar}{\Delta_R t}G^{2,0}_{1,3}\bigg(\frac{(\frac{\Delta_R}{\hbar}t)^2}{4}\Big\vert^{[],[3/2]}_{[0,1],[1/2]}\bigg)+{\color{blue}\frac{\pi}{2}\frac{\frac{\Gamma t}{\hbar}}{1+(\frac{\Gamma t}{\hbar})^2}}\bigg\}+(L\leftrightarrow R)+\frac{8}{\pi\zeta^2}\frac{d\delta(t)}{dt,}\label{Kntstsn}
\end{align}
where $H_j$ is the Struve function, and $G$ denotes the Meijer-G function. The terms marked in blue arise from the MZM mode. In the wide-band limit, the anomalous spectral function only contains an MZM contribution. In the general case with a finite bandwidth, the anomalous spectral function contains band contributions for $|\omega|\geq \Delta$, which would lead to corresponding terms in $K_S$ oscillating at a frequency commensurate with $\Delta/\hbar$, similar to those present in $K_N$. Nevertheless, since the quasiparticle bands do not produce NFJE, this assumption does not affect our analysis pertaining to the NFJE. 
\newpage
\section{Andreev Bound States in inhomogeneous BCS chain}
\begin{figure}[htb!]
  \includegraphics[height=4.4cm,width=.26\linewidth]{{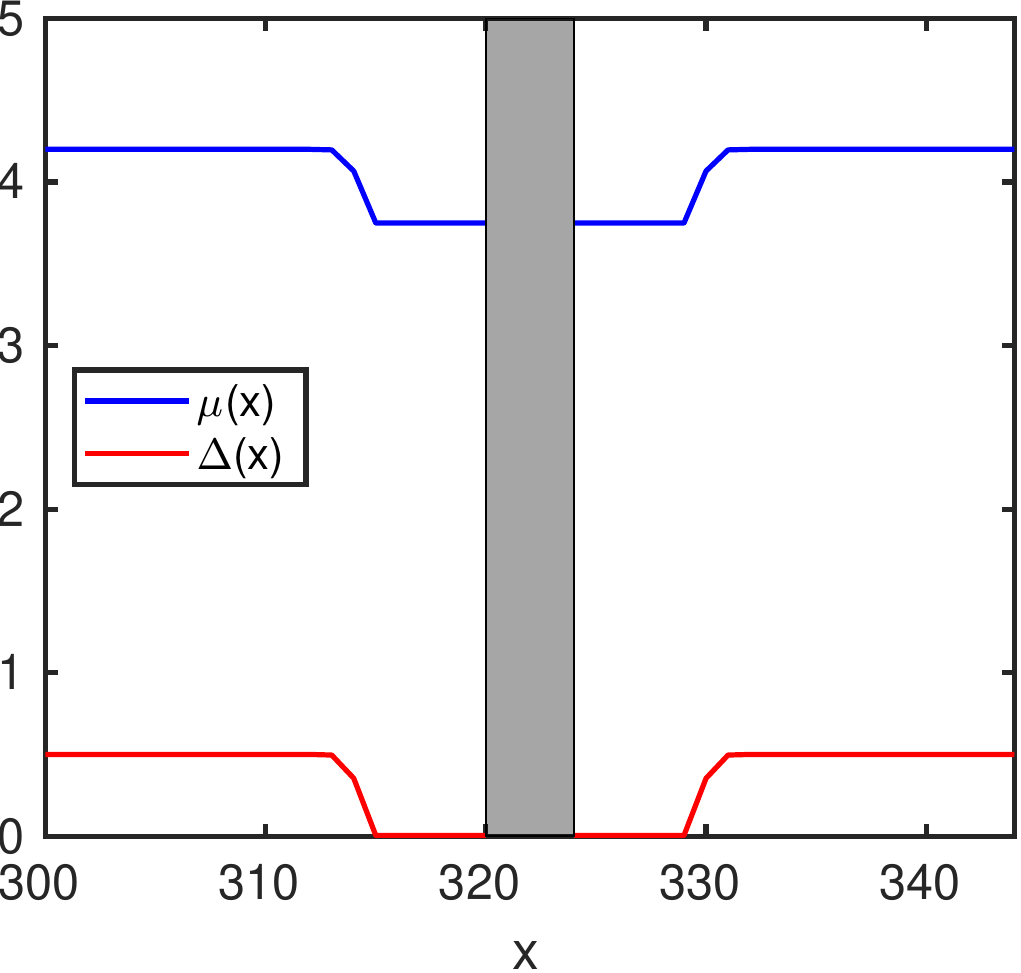}}%
\caption{The effective chemical potential and the pairing amplitude along the nanowire, with the tunnel junction marked by the thin shaded region. Each lead has $320$ sites in this example.}
\label{FigS3}
\end{figure}
In this section we consider an inhomogeneous s-wave superconducting nanowire with low-energy Andreev bound states developing due to inhomogeneity in the chemical potential $\mu$ and the pairing potential $\Delta$ near the junction, which in turn typically arise from gate-induced confinement potentials. We employ the Hamiltonian Eq. \eqref{Hsw}, but with the position-dependent chemical potential and pairing amplitude, as shown in Fig. \ref{FigS3}. We assume that both the leads are identical, apart from having the non-superconducting potential well on opposite ends (see Fig. \ref{FigS3} for a schematic).
\begin{figure}[htb!]
\includegraphics[height=5.2cm,width=.99\linewidth]{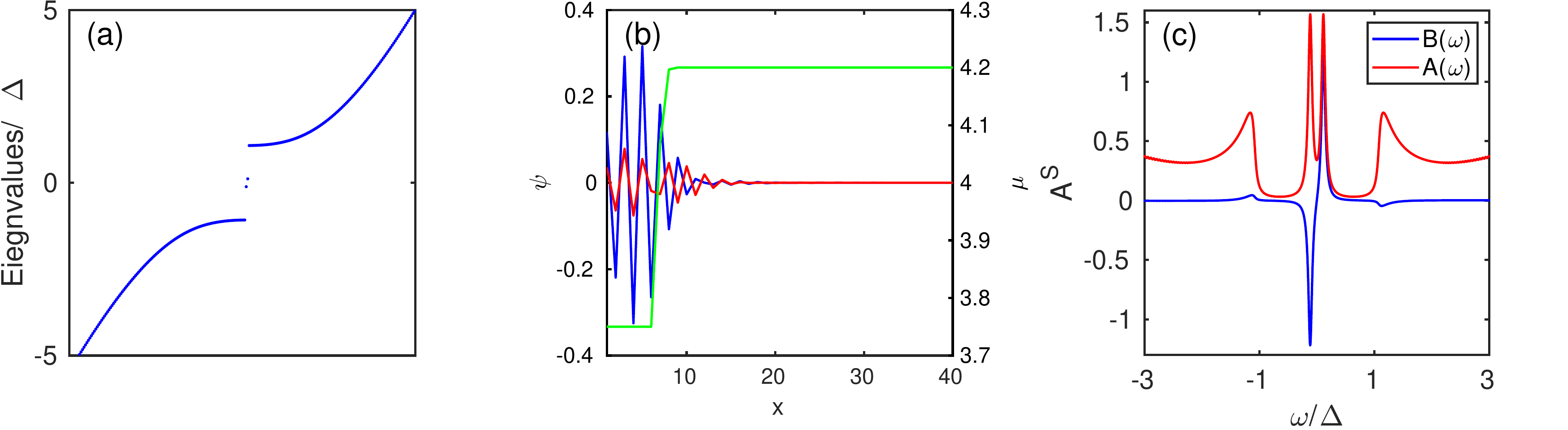}
\caption{(a) The eigenvalues (blue dots). Those lying below energy $\Delta$ are sub-gap ABS. (b) The ABS wavefunctions for the lead with a potential well on the left side, shown as a function of the lattice index along the length of the wire. The green curve, corresponding to the y-axis on the right side, shows the chemical potential profile. (c) The anomalous $B(\omega)$ and normal $A(\omega)$ surface spectral function, showing two low-energy ABS spectral peaks. We have considered $\Delta=0.5$, $\zeta=4$, $\mu=4.20$, $\Gamma=0.04\Delta$, with the chemical potential and pairing amplitude as shown in Fig.\ref{FigS3}.}
\label{FigS4}
\end{figure}

The results in this section can be readily extended to the case of partially separated Andreev bound states (psABS), which are near-zero-frequency Andreev bound states whose wavefunctions are partially separated, as opposed to being fully separated in the case of true MZMs, or fully overlapping in the case of trivial ABS. They arise in the presence of a non-superconducting smooth potential well at one end of the proximitised Rashba nanowire \cite{sKells2012,sLiu2017,sReeg2018,sMoore2018,sMoore2018a}, or infact, even in the simple case of a Kitaev chain. 
\newpage
\section{Response to Heaviside step voltage}
\label{stepJp}
In this section we calculate the response to the Heaviside step voltage $V(t)=V_0\Theta(t)$. From Eq. \eqref{ineqssf0}, we separate the normal and pair-currents,
\begin{align}
I(t)=&\underbrace{\frac{e\mathcal{T}^2}{\hbar}\Im  J_n(t)}_{I_N}+\underbrace{-\frac{e\mathcal{T}^2}{\hbar}\Im e^{-i\phi(t)} J_p(t)}_{I_S}=\frac{e\mathcal{T}^2}{\hbar}\Im J_N(t)+\frac{e\mathcal{T}^2}{\hbar}\big[\sin(\phi(t))\Re J_p(t)-\cos(\phi(t))\Im J_p(t)\big],
\end{align}
where we have defined,
\begin{subequations}
\begin{align}
J_n(t)=&\int_{0}^\infty d\tau e^{-i(\phi(t)-\phi(t-\tau))/2}K_N(\tau)\\
J_p(t)=&\int_{0}^\infty d\tau e^{i(\phi(t)-\phi(t-\tau))/2}K_S(\tau).
\end{align}
\end{subequations}
\begin{figure}[htb!]
\subfloat{
\stackunder[5pt]{ \includegraphics[height=5.2cm,width=.6\linewidth]{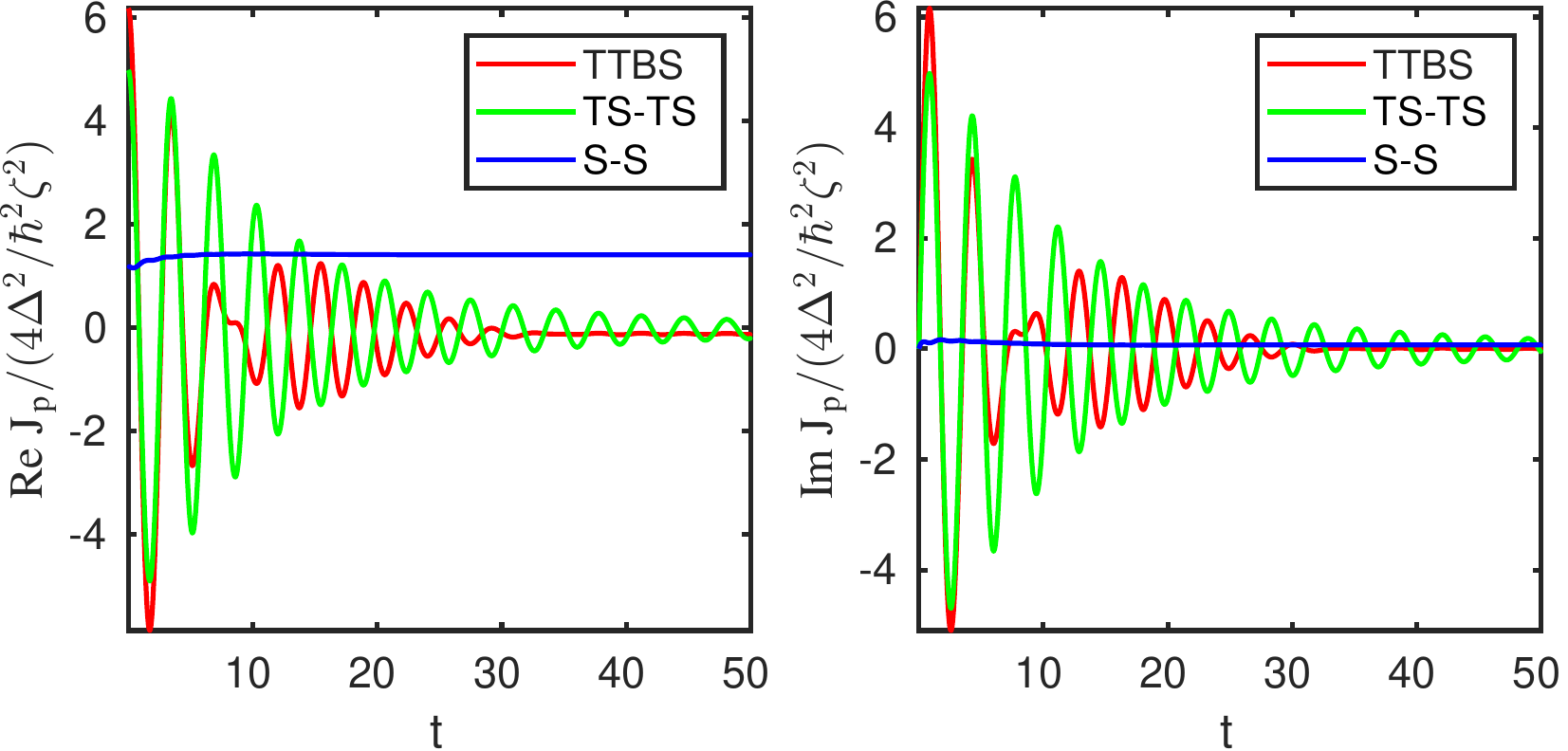}}{(I)}
}\hspace*{0.1cm}
\subfloat{
  \stackunder[5pt]{\includegraphics[height=5.4cm,width=.35\linewidth]{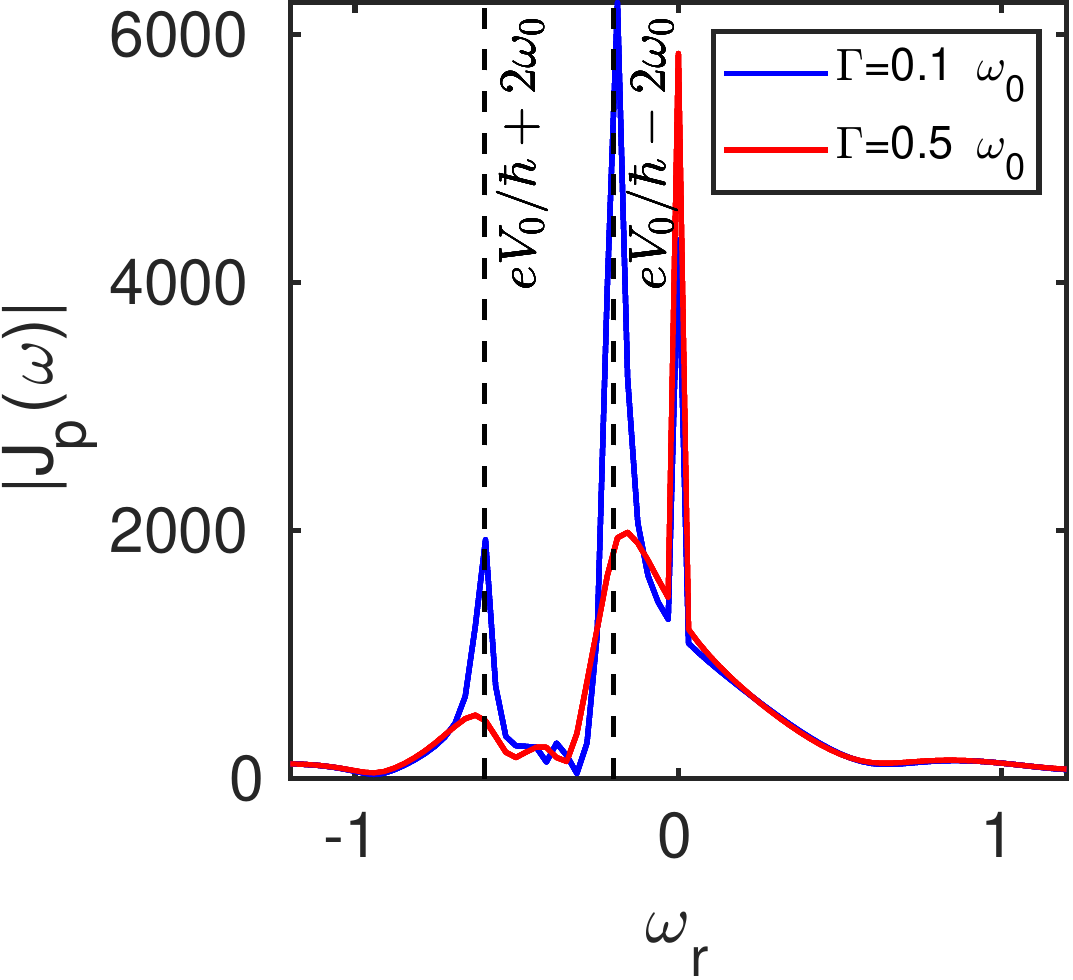}}{(II)}
}
\caption{(I) $J_p$, corresponding to a Heaviside-step voltage bias, with $eV=1.83\Delta$, $\zeta=150\Delta$, and $\Gamma=0.1\Delta$. In the S-S (two s-wave superconducting leads) case, $J_p$ is largely constant, with small modulations near $t=0$. In the S-S case with subgap bound states (TTBSs) (both leads having TTBSs at $\omega_0=0.1\Delta$, band contributions are neglected as they do not lead to NFJE), and the TS-TS case (topologically superconducting leads based on Kitaev chain), coherent $\omega_J/2-$oscillations are seen, eventually decaying over $t\sim \hbar/\Gamma$. (II) Fourier transform of $J_p(t)$ for TTBSs with $\omega_0=0.1\Delta$, $eV_0=0.4\Delta$. The two peaks at $eV_0/\hbar\pm 2\omega_0$ arise due to the combined dynamics of the bias voltage and the intrinsic time-evolution of the TTBSs, thereby defining the NFJE frequency. The response at the lower frequency $eV_0/\hbar- 2\omega_0$ dominates and primarily defines the time dependence. With increasing $\Gamma$, the NFJE decays faster, suppressing the peaks at $\omega_J/2\pm 2\omega_0$.}
\label{FigS5}
\end{figure}
Specialising to the case of the step voltage $V(t)=V_0\Theta(t)$, we obtain Eq. \eqref{iwerttimes} of the main text,
\begin{align}
I_S=&\frac{e\mathcal{T}^2}{\hbar}\bigg[\sin\bigg(\frac{2eV_0t}{\hbar}\bigg)\Re J_p(t)-\cos\bigg(\frac{2eV_0t}{\hbar}\bigg)\Im J_p(t)\bigg],
\end{align}
with,
\begin{align}
J_p(t)=e^{i\frac{eV_0t}{\hbar}}\int_t^\infty d\tau K_S(\tau)+\int_0^t d\tau e^{\frac{ieV_0\tau}{\hbar}}K_S(\tau).
\end{align}
\begin{figure}[htb!]
\subfloat{
\stackunder[5pt]{ \includegraphics[height=8.1cm,width=.44\linewidth]{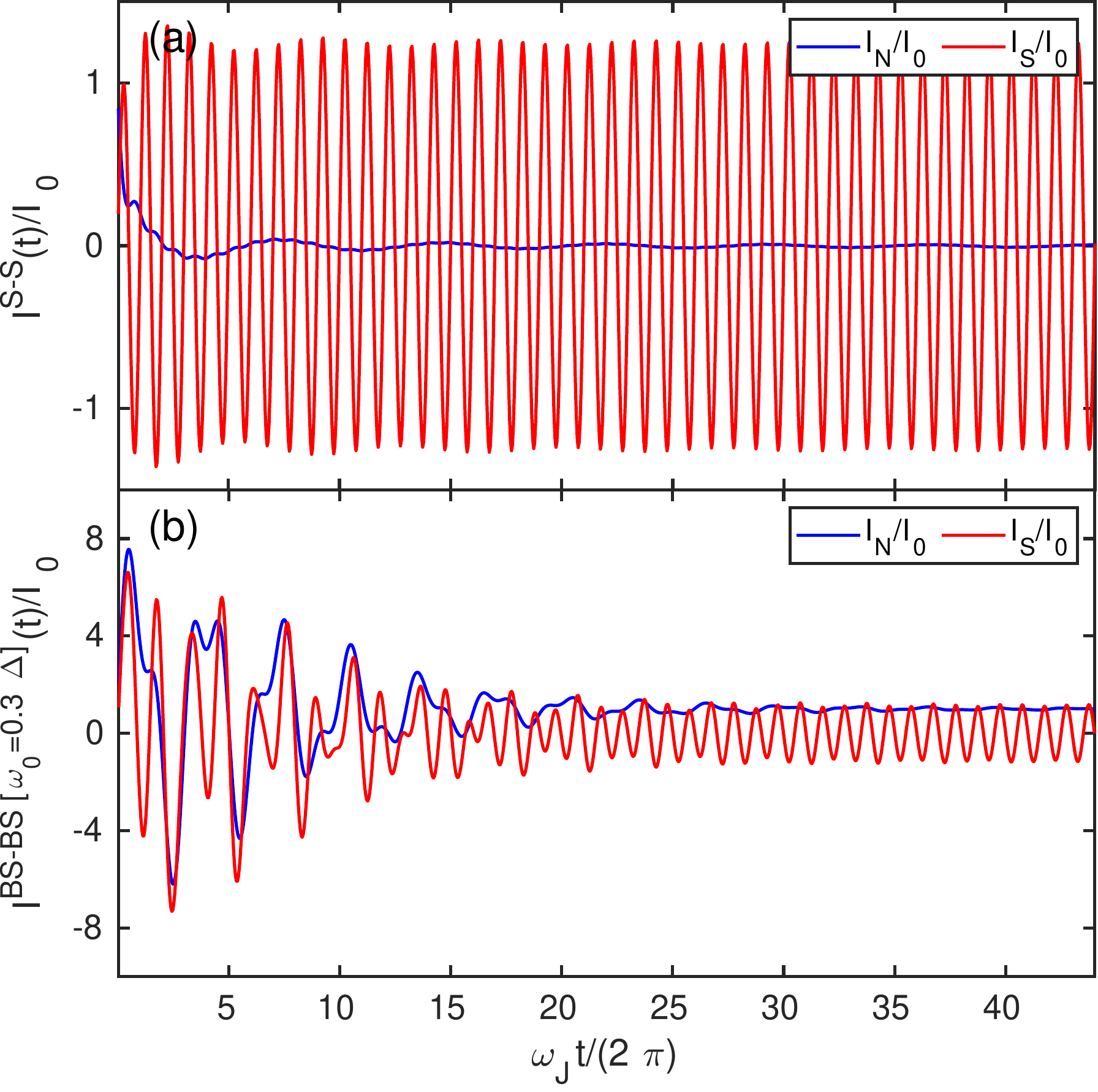}}{(I)}
}\hspace*{1cm}
\subfloat{
  \stackunder[5pt]{\includegraphics[height=4.5cm,width=.44\linewidth]{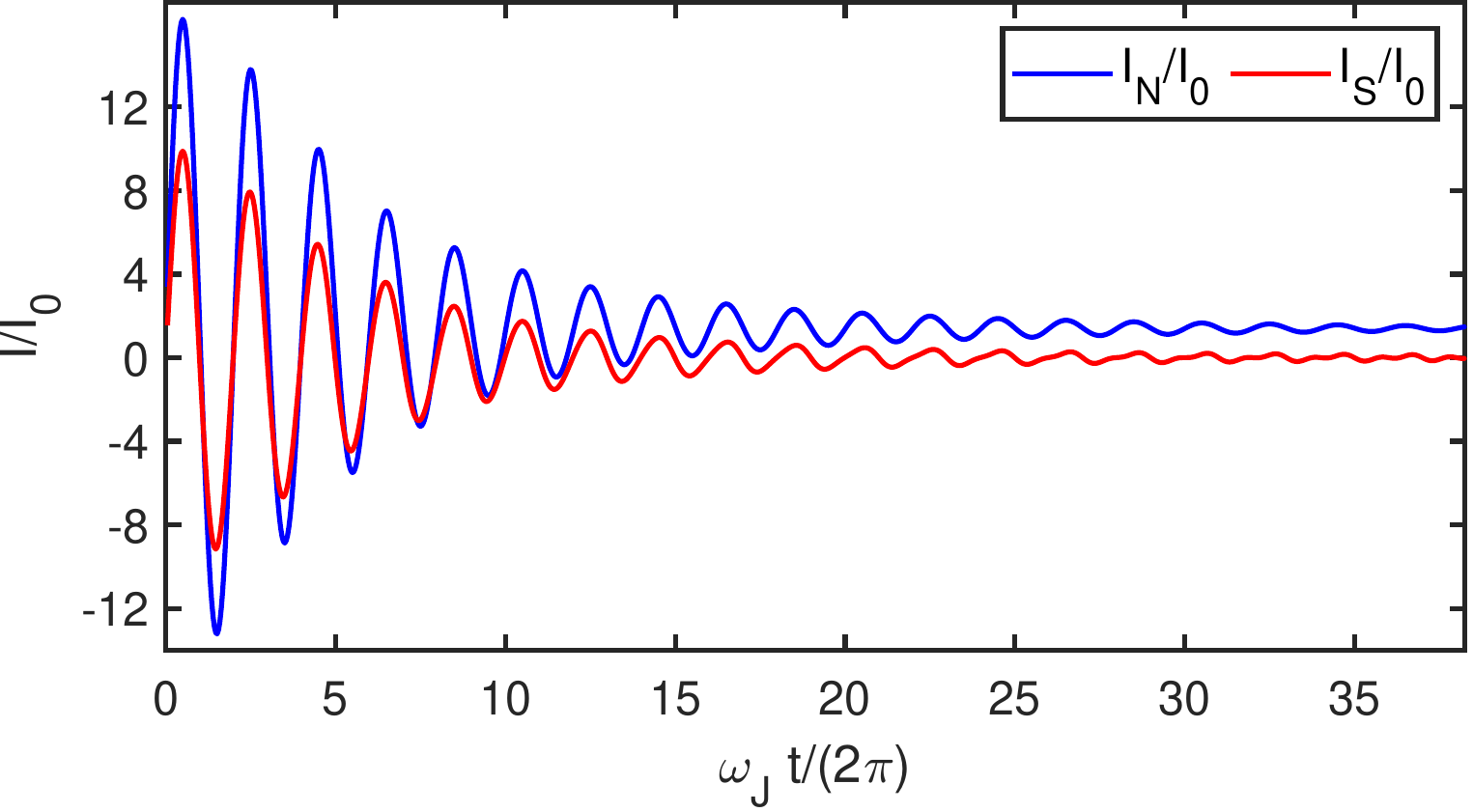}}{(II)}
}
\caption{Time-resolved current response to the Heaviside-step voltage bias. (I) For sub-gap bound states (TTBS), using same parameters and notation as in Fig. \ref{Fig2} in the main text, but with a larger $\omega_0=0.3\Delta$ for both leads. Non-bound state band contributions are neglected as they do not lead to NFJE. (I)(b) With the bound state located at a higher frequency $\omega_0=0.3\Delta$, the transient $\omega_J/2-$oscillations are significantly weaker as compared to the case with $\omega_0=0.1\Delta$ (Fig. \ref{Fig2}). (II)(b) The strongest transient $\omega_J/2-$oscillations are obtained with MZMs.}
\label{FigS6}
\end{figure}
\newpage
\subsection{Bound state}
For an analytical approximation, we consider the limit $eV_0\gg\omega_0\gg\Gamma$. In this limit, extracting the bound states contribution from the kernel in Eq. \eqref{Ksssabsn}, and considering $h_S^L=h_S^R=1$ for simplicity, we have,
\begin{align}
K_S(t)=&\frac{4\Delta_L\Delta_R}{\hbar^2\zeta^2}\frac{2\sin(2\omega_{0}t)}{\big(1+(\frac{\Gamma t}{\hbar})^2\big)^2}
\end{align}
Following the same procedure as before, we obtain,
\begin{align}
J_p(t)\approx&\frac{8\Delta_L\Delta_R}{\hbar^2\zeta^2}\frac{\hbar}{\Gamma}\Bigg\{-\frac{\frac{2\hbar\omega_0}{\Gamma}}{(\frac{eV_0}{\Gamma})^2-(\frac{2\hbar\omega_0}{\Gamma})^2}+e^{i t (\frac{eV_0}{\hbar}+2\omega_0)} \left(-\frac{1}{2 \left((\frac{\Gamma t}{\hbar})^2+1\right)^2\left(\frac{eV_0}{\Gamma}+\frac{2\hbar\omega_0}{\Gamma}\right)}+\frac{2 i \frac{\Gamma t}{\hbar}}{\left((\frac{\Gamma t}{\hbar})^2+1\right)^3 \left(\frac{eV_0}{\Gamma}+\frac{2\hbar\omega_0}{\Gamma}\right)^2}\right)\nonumber\\
&\hspace*{29.2mm}-e^{i t (\frac{eV_0}{\hbar}-2\omega_0)} \left(-\frac{1}{\left(2 \left((\frac{\Gamma t}{\hbar})^2+1\right)^2\right) \left(\frac{eV_0}{\Gamma}-\frac{2\hbar\omega_0}{\Gamma}\right)}+\frac{2 i \frac{\Gamma t}{\hbar}}{\left((\frac{\Gamma t}{\hbar})^2+1\right)^3 \left(\frac{eV_0}{\Gamma}-\frac{2\hbar\omega_0}{\Gamma}\right)^2}\right)\nonumber\\
&+e^{i\frac{eV_0}{\hbar}t}\Bigg[-\frac{\frac{\Gamma t}{\hbar} \sin (2 \omega_0 t)}{2\left((\frac{\Gamma t}{\hbar})^2+1\right)}+0.5 \left(\frac{2\hbar\omega_0}{\Gamma} \cosh\left(\frac{2\hbar\omega_0}{\Gamma}\right)-\sinh\left(\frac{2\hbar\omega_0}{\Gamma}\right)\right) \Re (\text{Ci}\left(2 \left(i+\frac{\Gamma t}{\hbar}\right)\frac{2\hbar\omega_0}{\Gamma})\right)\nonumber\\
&\hspace*{23.3mm}+0.5 \left(\cosh\left(\frac{2\hbar\omega_0}{\Gamma}\right)-\frac{2\hbar\omega_0}{\Gamma} \sinh\left(\frac{2\hbar\omega_0}{\Gamma}\right)\right) \Im(\text{Si}\left(2 \left(i+\frac{\Gamma t}{\hbar}\right)\frac{2\hbar\omega_0}{\Gamma})\right)\Bigg]\Bigg\}.
\end{align}
The first term, which contributes to the SJE, is clearly non-zero only for $\omega_0\neq 0$. Nevertheless, it is smaller than the remaining oscillating terms by atleast the factor $\hbar\omega_0/eV_0$, in the limit $eV_0\gg\hbar\omega_0\gg\Gamma$. The remaining terms oscillate at $\omega_J/2\pm2\omega_0$. Consequently, it follows from Eq. \eqref{iwerttimes} in the main text that $I_S$ oscillates at $\omega_J/2\pm2\omega_0$. Also, the prefactor in the expression for $J_p$ implies that the NFJE current has the amplitude $\sim(e\mathcal{T}^2/\hbar)(2\Delta_L\Delta_R/\hbar^2\zeta^2)(\hbar/\Gamma)=I_0(\Delta/\Gamma)$, as mentioned in the main text.

We present additional data for the response to the step voltage in Fig. \ref{FigS6}. In Fig. \ref{FigS6}(I) we see that the NFJE is weakened on using a larger value of $\omega_0=0.4\Delta$, as opposed to $\omega_0=0.1\Delta$ in Fig. \ref{Fig2}.

\subsubsection{Smooth steps}
\begin{figure}[!ht]
\subfloat{
  \stackunder[5pt]{\includegraphics[height=7.7cm,width=.49\linewidth]{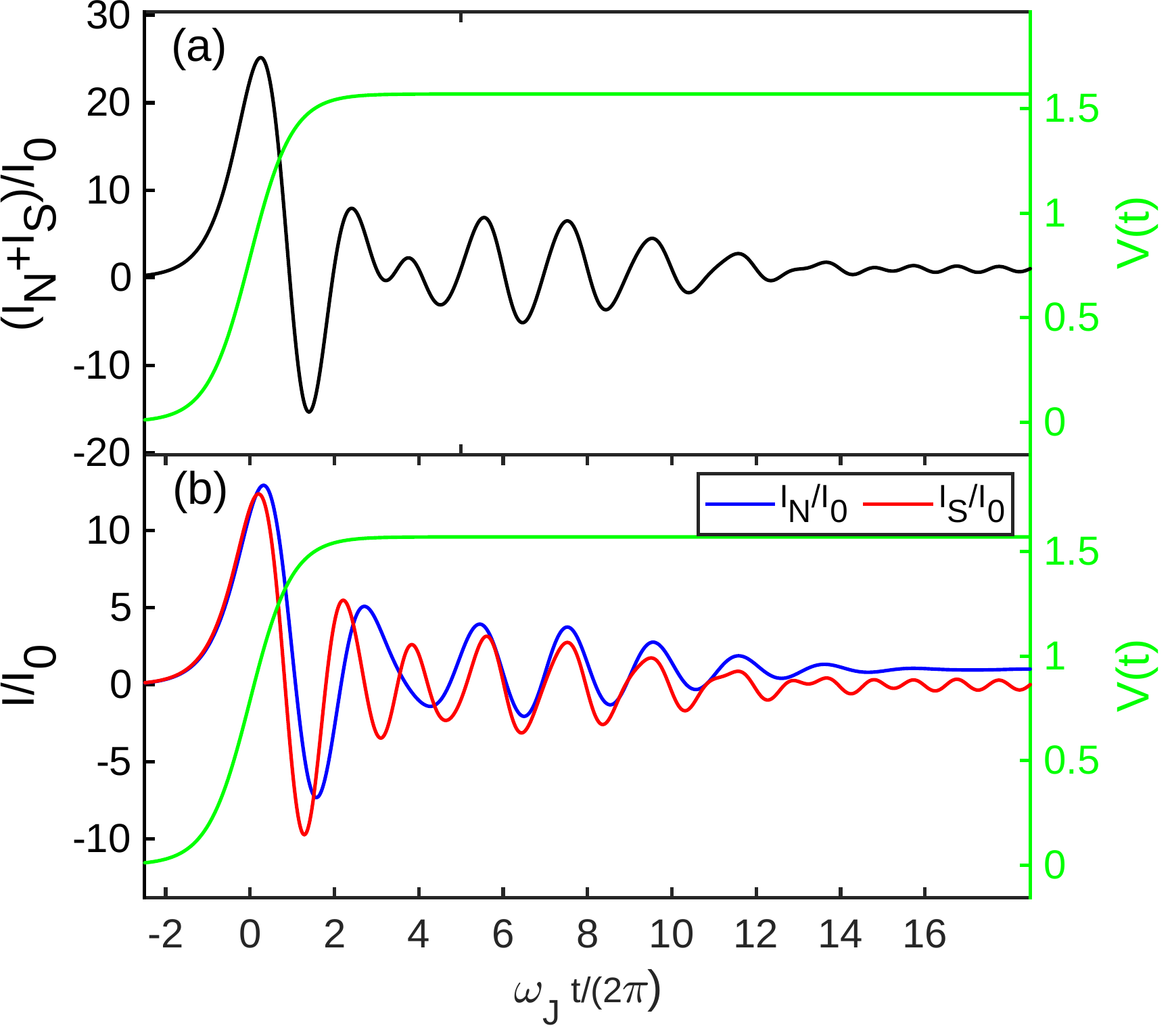}%
}{(I)}}\hspace*{5mm}
\subfloat{%
  \stackunder[5pt]{\includegraphics[height=7.7cm,width=.49\linewidth]{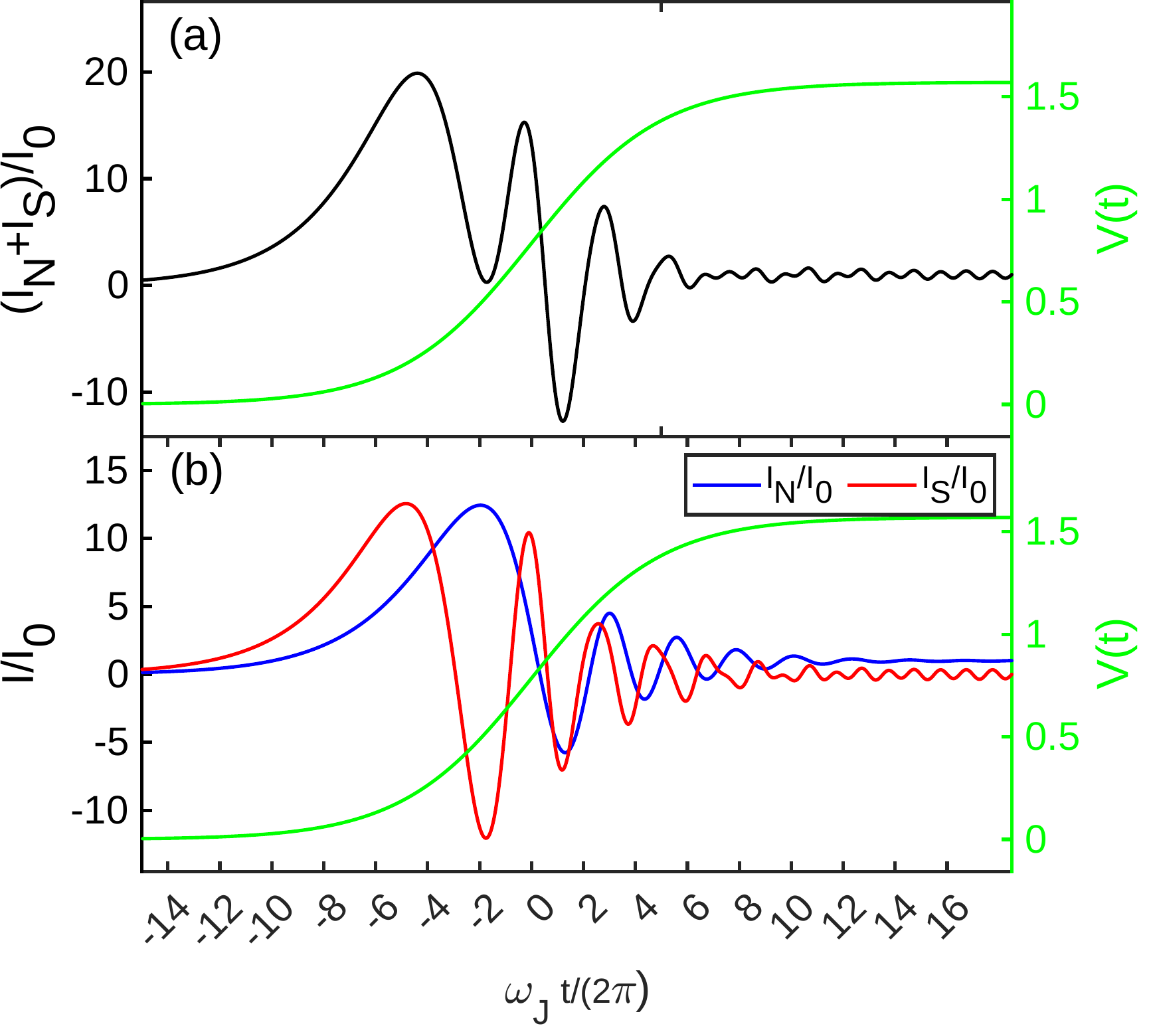}%
}{(II)}}
\caption{The response to a smooth heaviside step $V(t)=0.5(1+\tanh(t/\tau))$, with (I) $\tau=2$ and (II) $\tau=10$. We have considered the same parameters as in Fig.\ref{FigS2}. Note that $t^*=\hbar/\Gamma\equiv 1/\Gamma=10$.}
\label{FigS7}
\end{figure}

Here we consider the response to a smooth step, as mentioned in the introduction of the main text. During the rise of the step $t<\tau$, the oscillation frequency smoothly changes, which reflects the interference of waves tunneling at the present time $t$ and those having originated prior to the step. Correspondingly, they have the instantaneous oscillation frequency $\omega_J(t)/2 \approx eV(t<\tau)/\hbar$. Nevertheless, the SJE is recovered after $t^*=\hbar/\Gamma$. Additionally, the normal and pair currents tend to go out-of-phase with increasing $\tau$, which results in the NFJE oscillations decaying even sooner than $t^*$. This is evident from comparing Fig.\ref{FigS7} (I) and (II).

\subsection{MZM}
For TS leads hosting MZMs, using the spectral functions in the wideband limit for simplicity, only the MZM contributes to the anomalous spectral function and the kernel $K_S$. In this limit, we obtain the analytical result,
\begin{align}
J_p(t)=&\frac{4\Delta^2}{\hbar^2\zeta^2}\frac{\hbar}{\Gamma}\bigg\{\frac{1}{4} \frac{eV_0}{\Gamma} e^{\frac{eV_0}{\Gamma}} \bigg[-\text{Ei}\left(i \frac{eV_0 t}{\hbar}-\frac{eV_0}{\Gamma}\right)+\text{Ei}\left(-\frac{eV_0}{\Gamma}\right)+i \pi \bigg]-\frac{1}{4} \frac{eV_0}{\Gamma} e^{\frac{eV_0}{\Gamma}}\bigg[\text{Ei}\left(\frac{eV_0}{\Gamma}\right)-\text{Ei}\left(i \frac{eV_0 t}{\hbar}+\frac{eV_0}{\Gamma}\right)\bigg]+\frac{1}{2}\bigg\},\\
\xrightarrow{\Gamma\ll eV_0}&\frac{4\Delta^2}{\hbar^2\zeta^2}\frac{\hbar}{\Gamma}\bigg[i\frac{e^{i\frac{\omega_J}{2}t}}{2\big(1+(\frac{\Gamma t}{\hbar})^2\big)}\bigg]-i\bigg(\frac{\Gamma}{eV_0}\bigg)^2\bigg].\label{jpaproxstep}
\end{align}
The first term generates the NFJE, which has the magnitude $\sim I_0(\Delta/\Gamma)$. The second term is the SJE arising purely from the MZMs, and has the magnitude $\sim I_0(\Delta/\Gamma)(\Gamma/eV_0)^2$. In the generic case with a finite band-width, the band contributions also contribute to the anomalous spectral function and hence $J_p$. Hence, the SJE arising from the band bears the same magnitude as in the case of the S-S junction (see next section), $\sim I_0$, which is larger than the SJE arising purely from the MZMs.

We show the behaviour of $J_p$ in Fig. \ref{FigS5}, from which it is clear that sub-gap bound states and the MZMs have $\omega_J/2$ modulations, while the conventional S case remains largely feature-less. As in the case of bound states, the NFJE current has the amplitude $\sim I_0(\Delta/\Gamma)$.

In Fig. \ref{FigS6}(II), we consider the MZM, having $\omega_0=0$, which shows the strongest NFJE signal.
\subsection{S-S junction}
We focus on only the SJE here, as the NFJE response in the absence of any subgap states is negligible. In the steady state, 
\begin{align}
J_p=&\frac{4\Delta_L\Delta_R}{\hbar^2\zeta^2}\frac{\hbar}{\Delta}\times\begin{cases}
\frac{1}{\pi}K\left[\left(\frac{eV_0}{2\Delta}\right)^2\right];\quad |eV_0|<2\Delta\\
\frac{2}{\pi}K\left[\left(\frac{2\Delta}{eV_0}\right)^2\right]+i\frac{2}{\pi}K\left[1-\left(\frac{2\Delta}{eV_0}\right)^2\right];\quad |eV_0|\geq 2\Delta
\end{cases},
\end{align}
where $K(x)$ is the complete elliptic integral of the first kind. Hence, noting that $K(0)=\pi/2$, the $I_S$ has the amplitude $\sim(e\mathcal{T}^2/\hbar)(2\Delta_L\Delta_R/\hbar^2\zeta^2)(\hbar/\Delta)=I_0$, where $I_0$ is the standard critical current for s-wave superconductors.
\subsection{Andreev Bound States in inhomogeneous BCS chain}
\begin{figure}[htb!]
\includegraphics[height=8.6cm,width=.5\linewidth]{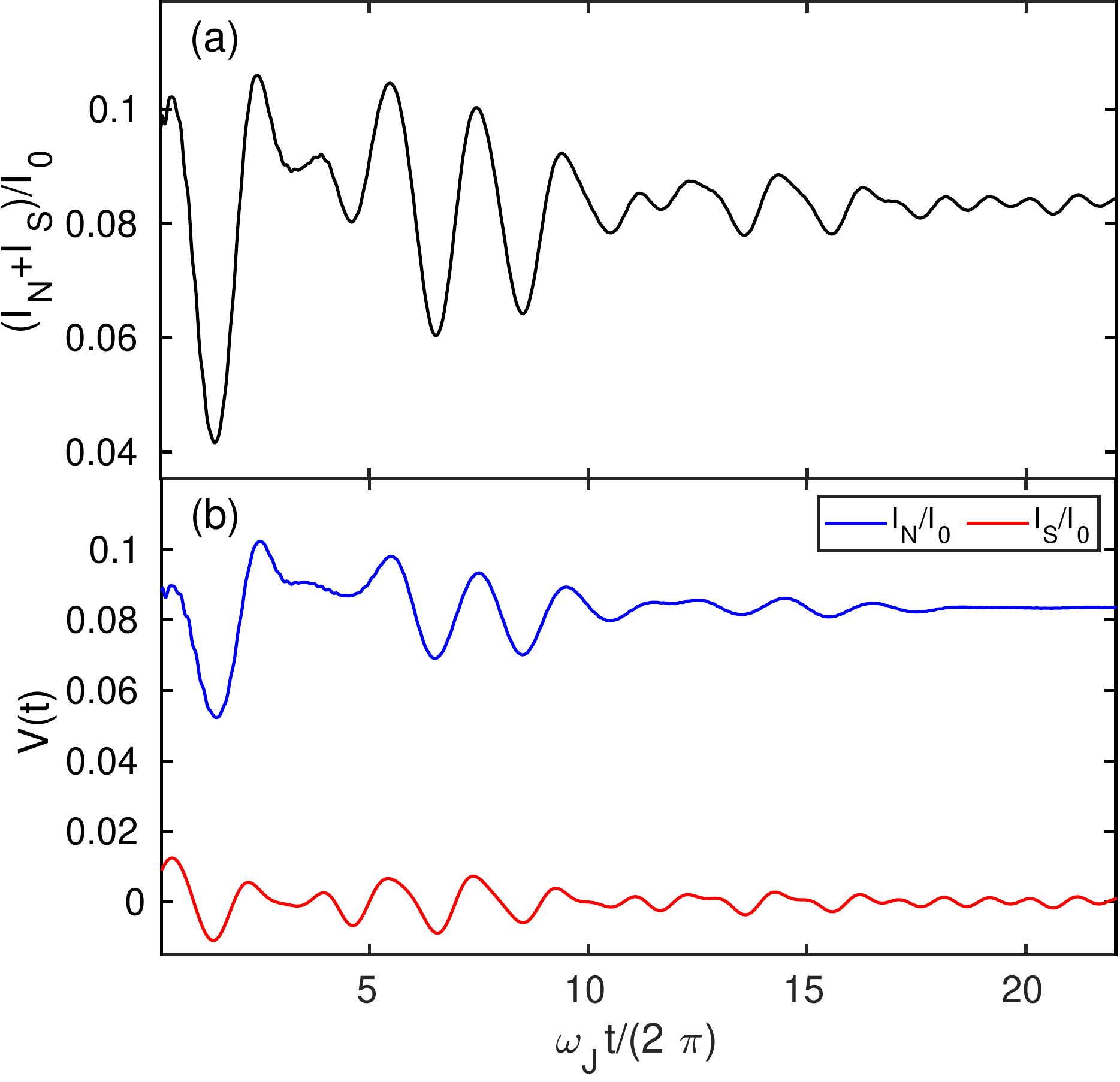}
\caption{The current response to a Heasviside step voltage. We have used the same parameters as in Fig. \ref{FigS4}.}
\label{FigS8}
\end{figure}

In Fig. \ref{FigS8} we consider the case of near zero-frequency Andreev bound states in inhomogeneous a BCS superconducting chain. Using the numerically obtained spectral functions (Fig. \ref{FigS4}(c)), we compute the kernels using Eq. \eqref{knksgen}, and subsequently the current using Eq. \eqref{ineqssf}. Finally, in Fig. \ref{FigS8}, where we show the response to the Heaviside step voltage $V(t)=V_0\Theta(t)$, we observe the transition from the initially dominant $\omega_J/2-$oscillations to the $\omega_J$ oscillations, as described in the main text.

\clearpage\section{Response to square wave voltage}
\subsection{Andreev Bound States in inhomogeneous BCS chain and Majorana zero modes}

\begin{figure}[!ht]
\subfloat{
  \stackunder[5pt]{\includegraphics[height=8.2cm,width=.5\linewidth]{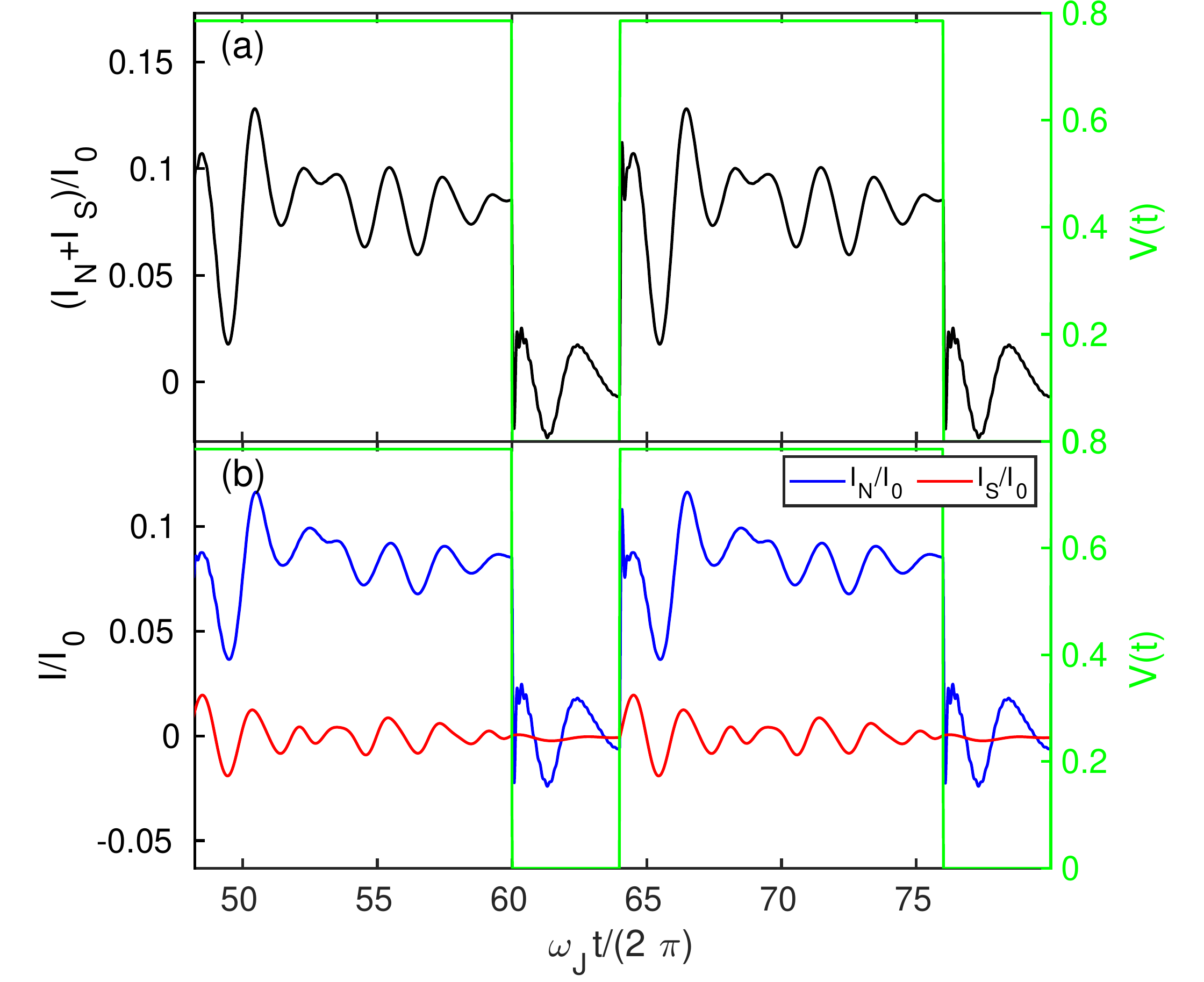}%
}{(I)}}\hspace*{5mm}
\subfloat{%
  \stackunder[5pt]{\includegraphics[height=5.4cm,width=.365\linewidth]{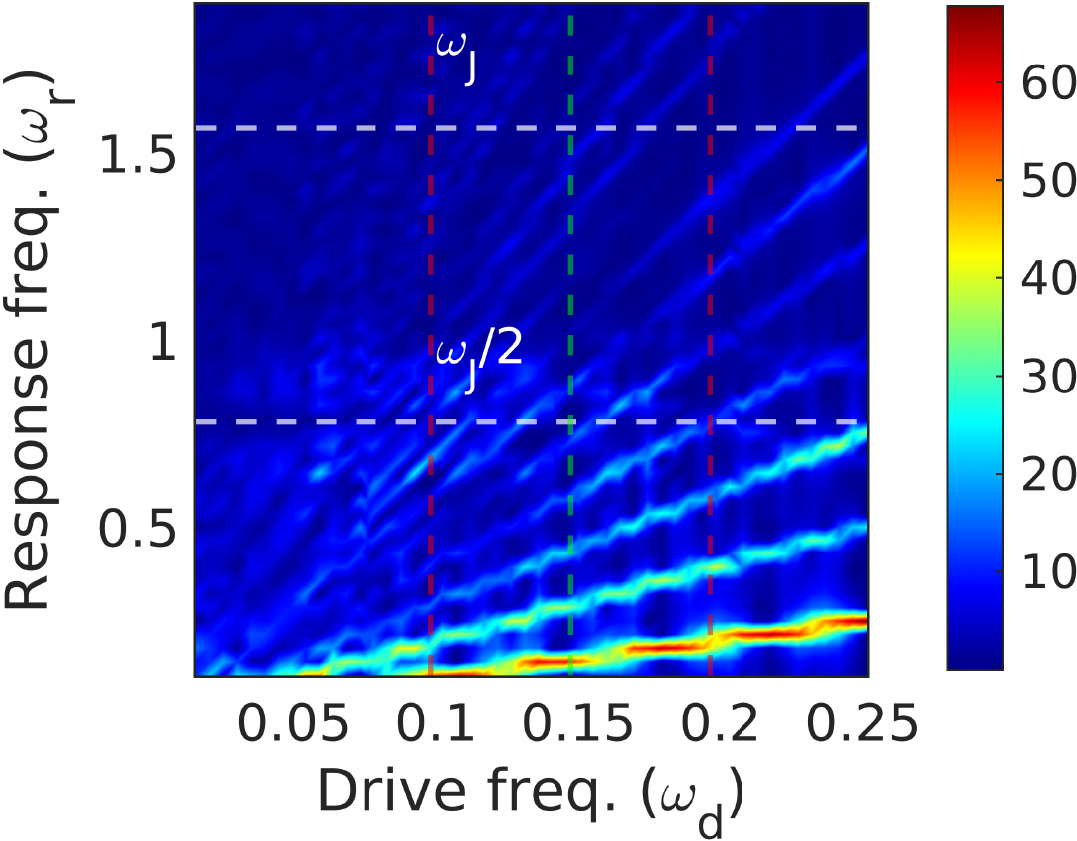}%
}{(II)}}\\
\subfloat{
  \stackunder[5pt]{\includegraphics[height=8.2cm,width=.5\linewidth]{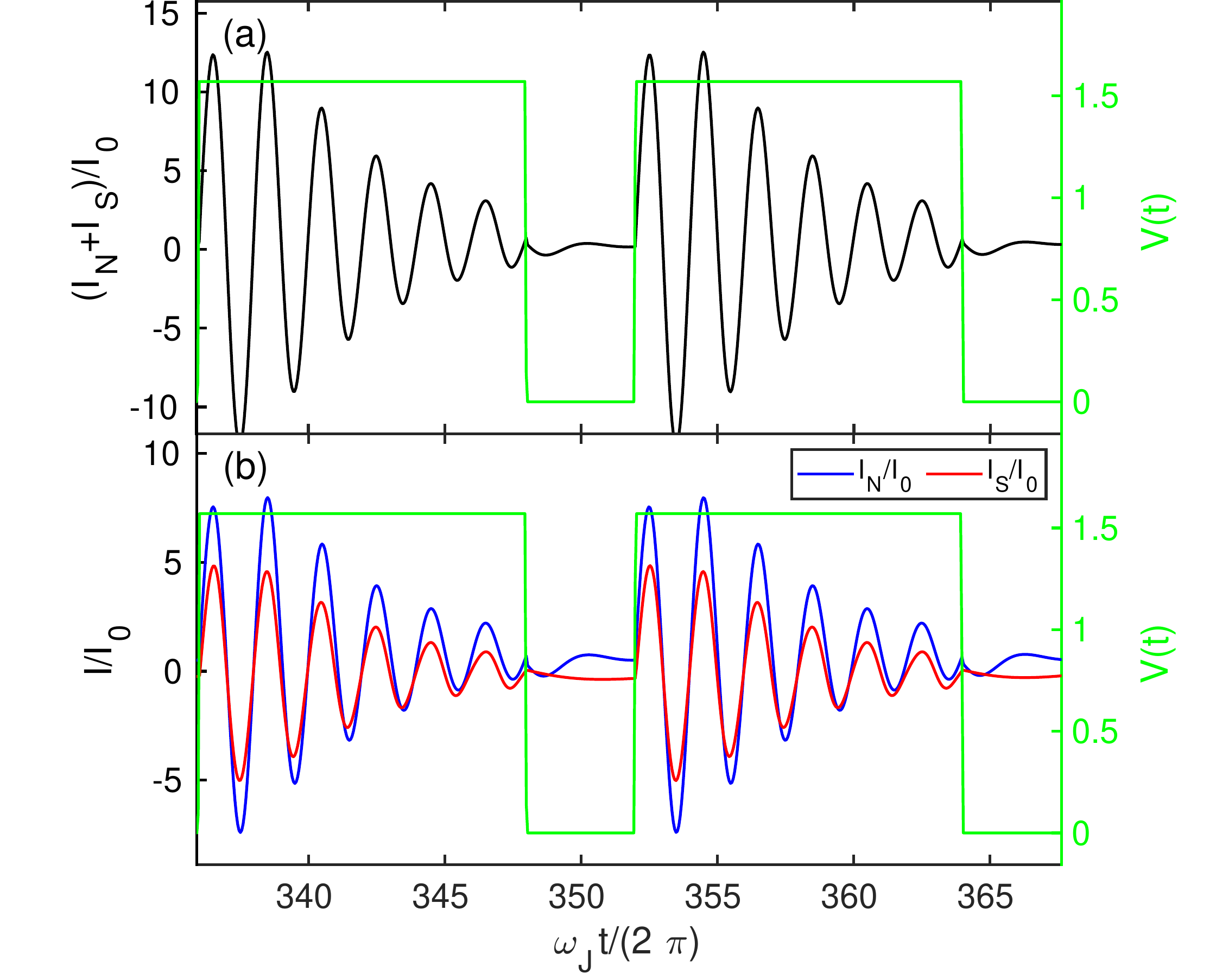}%
}{(III)}}\hspace*{5mm}
\subfloat{%
  \stackunder[5pt]{\includegraphics[height=5.4cm,width=.35\linewidth]{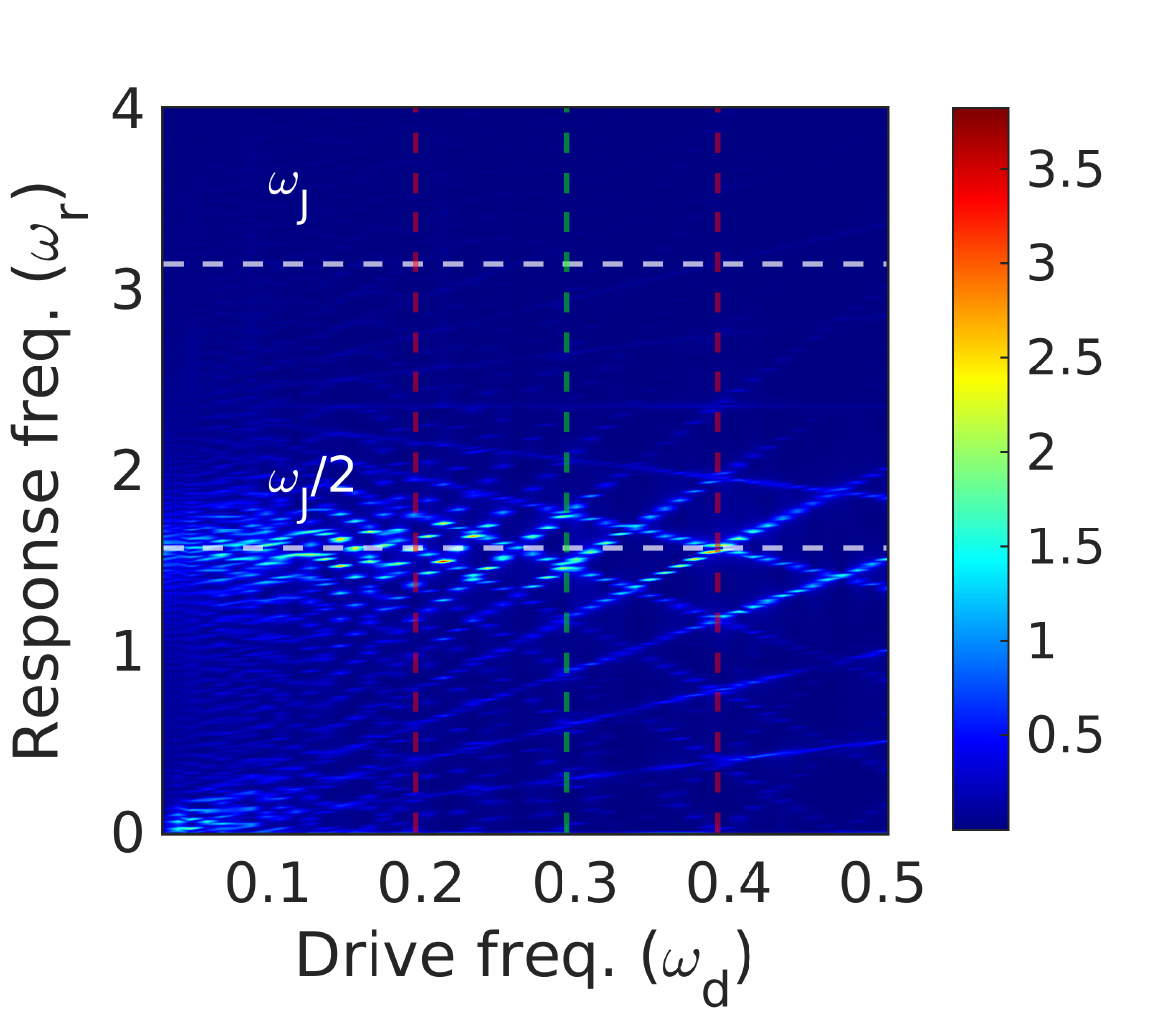}%
}{(IV)}}
\caption{(I) The response to the square wave for the S-S junction  low-energy ABS (same parameters as in Figs. \ref{FigS4} and \ref{FigS8}). We use $eV_0=(\pi/2)\Delta$, duty cycle $D=0.75$, and $\omega_d=(1-D)eV_0/2$ which ensures the strongest resonance. As in the main text, this combination ensures that the sections of the square pulses are commensurate with the NFJE oscillation period. The green curve shows the square wave bias (amplitude scaled to fit). (II) The Fourier spectra of the current reveals NFJE oscillations at frequency $\omega_r=\omega_J/2$. The low-frequency components at $\omega_d,2\omega_d\ldots$ arise from the response of the normal current to the bias, as evident from (I). (III) The response to the square wave for the TS-TS junction with MZMs, with $\Delta=1$ and $\Gamma=0.1\Delta$. We use the same $eV_0$ and $\omega_d$ as above. The green curve shows the square wave bias (amplitude scaled to fit). (IV) The Fourier spectra of the current in (III). The $\omega_J$ response is absent. }
\label{FigS9}
\end{figure}
\newpage
\section{Experimental distinction: Majorana zero modes vs low-energy topologically trivial bound states}
Here, we extend the comparison between non-topological near-zero-frequency TTBSs with $\omega_0\sim\Gamma$ and MZMs, by using a square-wave bias. Note that since $\omega_0\sim\Gamma$, it cannot be distinguished by conventional transport methods which have a resolution $\sim\Gamma$. Choosing the voltage $eV_0=4\omega_0$ as motivated in the main text, we obtain significantly distinct responses from the MZMs and TTBSs. While the former respond at the fractional Josephson frequency $\omega_{J,\text{MZM}}'=\omega_J/2$, the latter largely exhibit SJE at frequency $\omega_{J,\text{TTBS}}'=\omega_J$ as its NFJE is extremely short-lived and more importantly, inconspicuous, due to our choice of voltage $V_0$.
\begin{figure}[!ht]
\includegraphics[height=8.0cm,width=.8\linewidth]{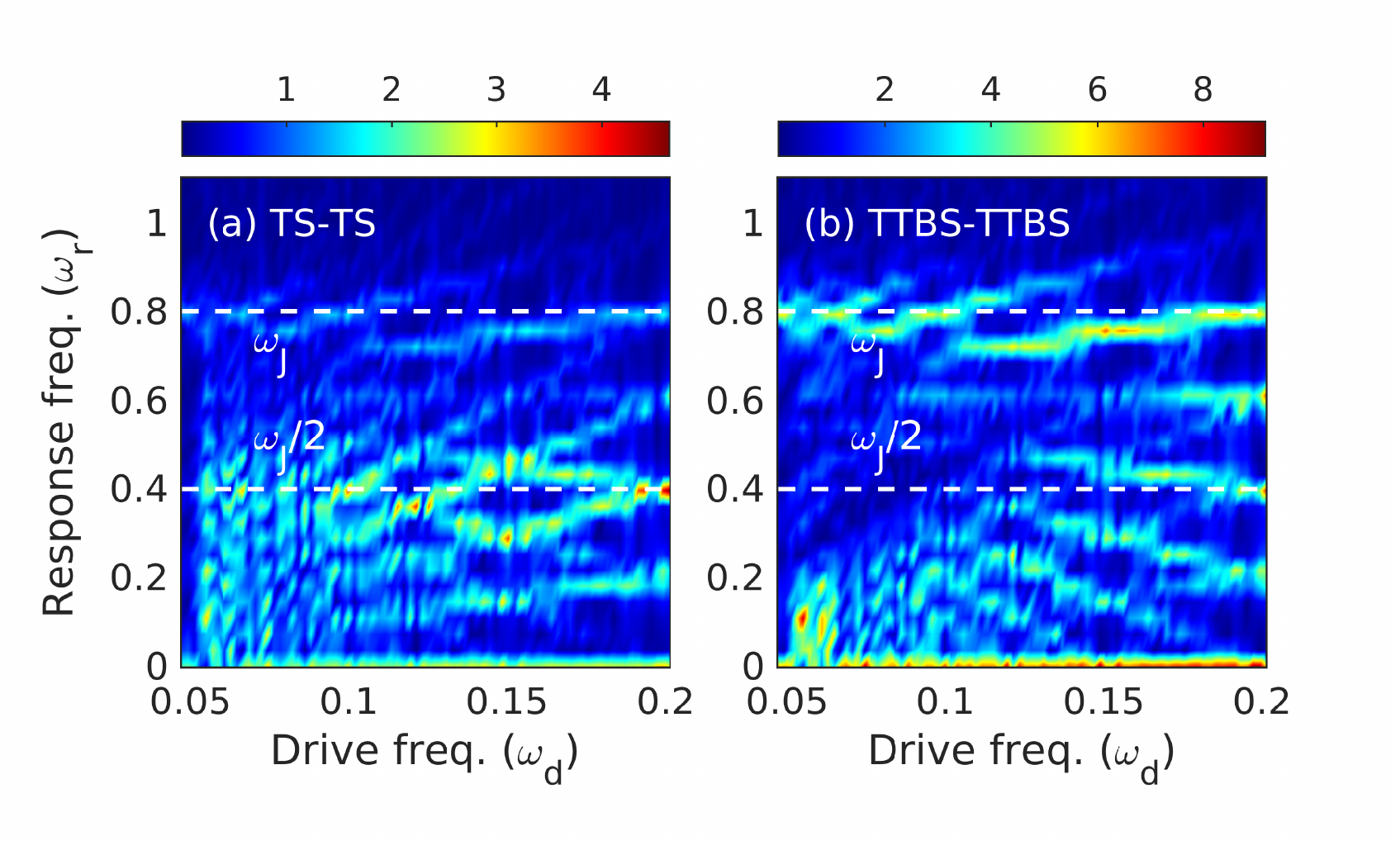}
\caption{Following the discussion associated with Fig. \ref{Fig3} in the main text, using the same parameters, we extend it to a square-wave bias. We show the normalised current $(I/I_0)$ resolved in Fourier domain, with varying drive frequency $\omega_d$. The fractional $\omega_J/2$ response dominates in the TS case, while the TTBS largely shows SJE. }
\label{FigS10}
\end{figure}

\FloatBarrier


\begin{thebibliography}{70}
\bibitem{bandneg} We only consider the TTBS in the spectral function as the band state contributions rapidly decohere due to the mutual interference of IQPs with a large range of energies. As such, they do not qualitatively affect the NFJE, and are thus neglected here. The cases with topologically superconducting leads, however, include both MZM and band contributions. We elaborate on a concrete case with BCS superconducting leads considering both bound and band states in the SM~\cite{sm}.
\end{thebibliography}
\end{document}